\documentclass[superscriptaddress,showpacs,showkeys,nofootinbib,reprint]{revtex4}
\usepackage{epsfig}
\usepackage{amsmath}

\usepackage[english]{babel}       

\newcommand{\ba}{\begin{array}{ccc}}
\newcommand{\ea}{\end{array}}

\newcommand{\bg}{\begin{equation}}
\newcommand{\beq}{\begin{equation}}

\newcommand{\ee}{\end{equation}}
\newcommand{\eeq}{\end{equation}}
\newcommand{\bgs}{\begin{eqnarray}}
\newcommand{\bea}{\begin{eqnarray}}
\newcommand{\ens}{\end{eqnarray}}
\newcommand{\eea}{\end{eqnarray}}

\newcommand{\lp}{\left(}
\newcommand{\rp}{\right)}

\begin{document}

\title {Feynman Diagrams and Rooted Maps}

\author{A. Prunotto}
\affiliation{Dipartimento di Fisica dell'Universit\`a di Torino and \\ 
  Istituto Nazionale di Fisica Nucleare, Sezione di Torino, \\ 
  via P.Giuria 1, I-10125 Torino, Italy}

\author{W.M. Alberico}
\affiliation{Dipartimento di Fisica dell'Universit\`a di Torino and \\ 
  Istituto Nazionale di Fisica Nucleare, Sezione di Torino, \\ 
  via P.Giuria 1, I-10125 Torino, Italy}

\author{P. Czerski}
  \email{czerski@up.krakow.pl}
\affiliation{Institute of Computer Science, The Pedagogical University of Cracow,\\
     ul.Podchor\c a\.zych 2, PL-30-084 Krak\'ow, Poland}

\begin{abstract}
The  Rooted Maps Theory, a branch of the Theory of Homology, is shown to be 
a powerful tool for investigating  the topological properties of Feynman 
diagrams, related to the single particle propagator in the quantum many-body 
systems. The numerical correspondence between the number of this class of 
Feynman diagrams as a function of perturbative order and the number of rooted 
maps as a function of the number of edges is studied. A graphical procedure to 
associate Feynman diagrams and rooted maps is then stated. Finally, starting 
from rooted maps principles, an original definition of the genus of a 
Feynman diagram, which totally differs from the usual one, is given.
\end{abstract}


\maketitle

{\it The problem of a correct and convenient counting of connected Feynman diagrams was often raised by Alfredo Molinari during his lectures in many-body physics. This paper took origin from his inquires on the subject and is dedicated to his memory.}

\section{Introduction}
Important progress has been made in recent years in developing the interplay between theoretical 
physics and graph theory (Floer \cite{floer1,floer2,floer3}, Fukaya 
\cite{fukaya1,fukaya2,fukaya3,fukaya4}, Schaeffer 
\cite{sch1,sch2,sch3,sch4,sch5,sch6}). Mathematicians and physicists worked to 
combine theories of planar trees and rooted maps with the enumeration of Feynman 
diagrams for field theories, essentially quantum electrodynamics (QED), quantum 
gravity and quantum computing (Atiyah \cite{ati}, Witten \cite{witt1,witt2}, 
t'Hooft \cite{thooft}, De Wolf \cite{dewolf1}, Di Francesco \cite{DiFr}). 
The problem of counting Goldstone Diagrams has already been solved,  Rossky
and Karplus \cite{R&K}, but it is a rather difficult task to 
list all publications about Feynman diagrams counting. Here we first focus the 
attention on the combinatorial point of view studied in depth by Kucinskii and 
Sadovskii \cite{Ku}. A graphical approach was followed for instance by Kleinert, 
Pelster, Kastening and Bachmann \cite{Kl}. A more physical approach was followed 
by Riddell \cite{Ridd} and then by Brouder \cite{BrI,BrII,BrIII}. 
A strictly theoretical physics point of view in this field was explored instead by 
Cvitanovi\'c, Lautrup and Pearson \cite{C}, with conclusions identical to the ones
of Arqu\`es and B\'eraud \cite{Arques-rooted,Arques-torus}. These
 results will be discussed in the following Sections.
 
 In particular we shall present an original method to transform a Feynman diagram of the 
 perturbative series expressing the single-particle propagator in many-body theory into a 
 well defined rooted map. At variance with quantum electrodynamics, where Furry's 
 theorem \cite{furry} entails the cancellation of certain classes of diagrams, in the 
 many-body approach the only cancellation occurring concerns the so-called disconnected 
 diagrams: indeed in the many-body problem  every connected 
diagram makes its contribution to the total amplitude (or Green's function). 
This renders the counting of Feynman diagrams somewhat more difficult. 

In last years, it has  been discovered 
that the {\em number} of Feynman diagrams with regard to the perturbative order 
and the {\em number} of rooted maps as a function of edges (and regardless to 
genus and number of vertices) is the same. The strength of the numerical  
relation between Feynman diagrams and maps may suggest important links between 
general relativity and quantum mechanics, providing we are able to trace, 
through the shape of the Feynman diagram, the topological properties related to 
its opposing party rooted map.

The analytical and/or numerical evaluation of higher order terms in a perturbative series 
becomes more and more complicated: thus one could come to the conclusion that the exact 
enumeration of the corresponding Feynman diagrams is not so useful. Facing this doubt we would 
like to cite Cvitanovi\'c, Lautrup and Pearson \cite{C}: ``In trying to 
understand the behavior of field theory at  large orders in perturbation theory, 
one finds that the number of diagrams contributing is an important effect. It is 
the cause of the combinatorial growth of amplitudes for superrenormalizable 
theories.''  Hence we will proceed in the major task of 
this article, which is the investigation of a rule to enumerate Feynman 
many-body diagrams at various perturbative orders.

The paper is organized as follows: in Section \ref{sec2} we shall briefly recall the main definitions 
and properties of topological maps, including the various enumerations of rooted maps (with or 
without taking into account specific characteristics of them). In Section \ref{sec3}, after a very short 
reckoning of the many-body single-particle Green's function and its perturbative expression, we 
shall make explicit the connection between the corresponding Feynman diagrams and the rooted 
maps obtained from them. Moreover the topological properties of the considered Feynman diagrams, 
in particular the genus, will be discussed. Finally Section \ref{sec4} will present our conclusions, leaving 
original details concerning the construction of the third (and higher) order
Feynman diagrams to the Appendix.

\section{Topological maps and rooted maps}\label{sec2}

 A {\em map} is, roughly speaking, a partition of a closed, connected 
two-dimensional surface into simply connected polygonal regions by means of a 
finite number of simple curves (or edges) connecting pairs of points called 
vertices in such a way that the curves are disjoint from one another and from 
the vertices. The enumeration of {\em planar maps} (maps on the sphere or on the 
projective plane) has been extensively investigated since 1960, in particular by 
W. T. Tutte \cite{Tutte1,Tutte2,Tutte3}. W. G. Brown then counted several types 
of maps on the projective plane and began investigating the torus, but did not 
obtain an explicit formula for counting maps on the torus 
\cite{Brown1,Brown2,Brown3}.

Walsh and Lehman \cite{WI,WII,WIII} presented the first census of maps on {\em 
oriented surfaces} of arbitrary genus in the early 70's. A map was defined to be 
a {\em rooted} one (see section \ref{def}) if one edge is distinguished, 
oriented and assigned a left and a right side. But since these authors worked on 
oriented (or, equivalently, orientable) surfaces, it suffices to distinguish and 
orient one edge-end because its left and right side are determined by the 
orientation of the surface. They considered 
two maps to be equivalent if they are related by an orientation preserving 
homeomorphism, according to Cairns \cite{Cai} and Tutte \cite{Tutte2}, which 
leaves fixed the distinguished edge-end (called 
the {\em root}). We will use this work in order to show a graphical 
correspondence between maps and Feynman diagrams in the realm of quantum 
many-body theory.

At the end of the past century, Arqu\`es and B\'eraud 
\cite{Arques-rooted,Arques-torus} presented a different approach in the study of 
maps: they enumerated rooted maps without regard to genus -- the genus of a map 
is essentially the genus of the embedding surface. They showed the existence of 
a new type of equation for the generating series of these maps enumerated with 
respect to edges and vertices: the Riccati's equation. By means of Riccati's 
equation Arque\`s and B\'eraud obtained a differential equation for the 
generating series of rooted trees regardless of the genus and as a function of 
edges which leads to a continued fraction for the generating series of rooted 
genus-independent trees and to a beautiful, unexpected relation between both 
previous generating series of trees and rooted maps. We will show that there 
exists a one to one relation between the number of rooted maps on orientable surfaces 
regardless to genus and with respect to edges and the number of Feynman diagrams 
for the one particle exact propagator as a function of perturbation order.

As we pointed out before, Walsh and Lehman \cite{WI,WII,WIII} studied 
enumeration of rooted maps by genus, edges and vertices. Bender, Canfield and 
Robinson \cite{BI,BII,BIII,BIV} studied in depth rooted maps on the torus and on 
the projective plane. 
\begin{figure}
\includegraphics[width=0.4\textwidth]{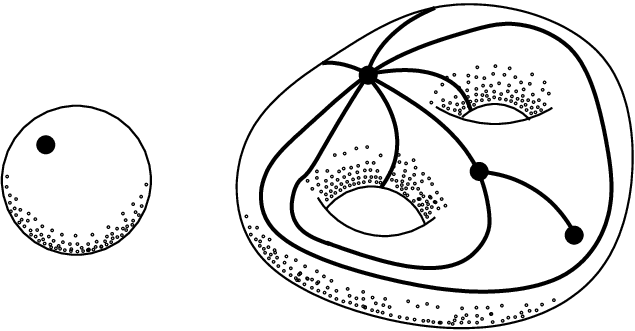}
\caption{Two examples of topological maps on orientable surfaces: left, the 
easiest topological map (single vertex on a sphere, no edges and one face); 
right, an example of a map on an orientable surface of genus $2$, familiarly 
called a \emph{cow}. In this second map we can locate three vertices, six edges 
and one face.} \label{2holes}
\end{figure}
Moreover, many years ago J. Touchard \cite{Touch} studied a problem of geometric configuration that 
actually corresponds to the enumeration on rooted maps, even though at the end of his work there 
seems to be a contradictory result. Work is presently going on on rooted maps (see for example 
 Courcelle and Dussaux \cite{Cou}, Krikun and Malyshev \cite{KI,KII,M}, Shaeffer and 
 Poulalhon \cite{SI,SII,SIII}, Jackson and Visentin \cite{J}, Yanpei \cite{Y}).

\subsection{Definitions} \label{def}
\subsubsection{Topological maps}

A topological map $M$ (see Fig.\ref{2holes}) on an orientable 
surface\footnote{A regular surface is orientable if one can give an orientation 
on it; roughly speaking, a regular surface $S\subset \mathbf{R}^n$ is denoted as 
an {\em orientable surface} if each tangent space $T_u$ in $u \in S$ can be 
connected with any other $T_u$ with a continuous function preserving the 
orientation of $T_u$.} $\Sigma \subset \mathbf{R}^3$ is a partition of $\Sigma$ 
in three sets:

\begin{figure}
\begin{minipage}[b]{0.47\textwidth}
\centering
\includegraphics[scale=.7]{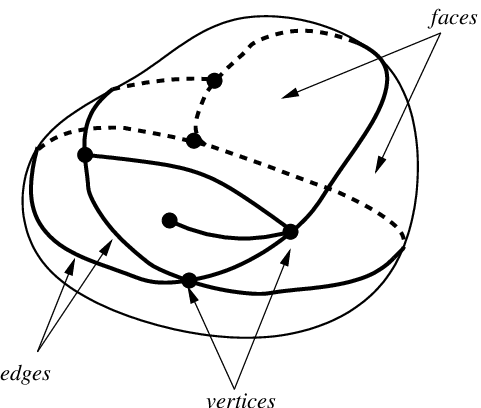}
\caption{Example of a topological map on an orientable surface.}
\label{map}
\end{minipage}
\hspace{0.5cm}
\begin{minipage}[b]{0.47\textwidth}
\centering
\includegraphics[scale=.7]{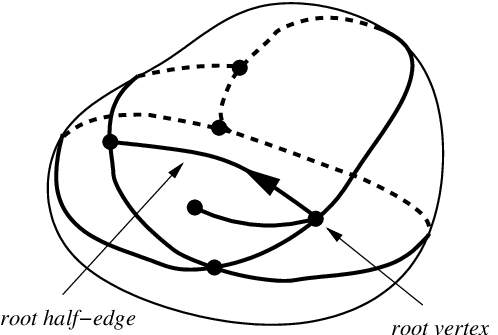} 
\caption{Example of a rooted map on an orientable surface of genus $0$.}
\label{rooted}
\end{minipage}
\end{figure}

\begin{itemize}
\item A finite set of points of $\Sigma$, called the {\em vertices} of $M$; 
\item A finite set of simple open Jordan arcs\footnote{A Jordan arc is an arc 
homeomorphic to a straight line segment.} lying on $\Sigma$, disjoint in 
pairs\footnote{If a Jordan arc contains one or more vertices inside, it splits 
into two or more edges.}, whose extremes are vertices, denoted as the {\em 
edges} of the map; 
\item A finite set of {\em faces}. Each face is homeomorphic\footnote{An 
homeomorphism is a bijective and bicontinuous function connecting each point of 
$\Sigma$ with each one of $\Sigma^{\prime}$; it ensures that starting object 
topology is kept.} to an open disc. Its border is the union of vertices and 
edges.
\end{itemize}
Obviously we may build an orientation on $\Sigma$ since we have chosen the 
surface $\Sigma$ to be orientable. An oriented edge of the map is a {\em 
half-edge}. Evidently to each half-edge are associated its starting vertex and 
its ending vertex (and the underlying edge). The  {\em genus} of the map $M$ is 
the genus\footnote{The genus of a surface is, in a simple counting, the number of its 
holes: e.g. the three-sphere has genus $0$, the torus has genus $1$ and so on.} 
of the host surface (also known as {\em embedding surface}).
\subsubsection{Rooted maps}
A map is called a {\em rooted map} if a certain half-edge is specified among the 
set of the half-edges. This peculiar half-edge, becomes the {\em root half-edge} 
of the map. In short, the initial vertex is the {\em root} of the map and we can 
refer to it as the origin of the map. Here we meet the first important analogy 
with Feynman diagrams: there exists an ``initial vertex'', which corresponds, in 
quantum many-body theory, to the initial point in space-time whence we start to 
calculate the propagator. See Fig.\ref{map} and Fig.\ref{rooted}.

The mathematical definition of rooted map is a little more subtle. Actually, two 
rooted maps with the same genus, associated with two surfaces $\Sigma$ and 
$\Sigma^{\prime}$  are {\bf isomorphic} -- i.e. can be regarded as a single 
rooted map -- if the following conditions are satisfied:
\begin{itemize}
\item There exists a {\em homeomorphism} between the two surfaces $\Sigma$ and 
$\Sigma^{\prime}$;
\item This homeomorphism preserves the orientation of the surface;
\item It maps vertices, edges, faces and the root half-edge of the first map 
into the homologous elements of the second one.
\end{itemize}
In fact it is the entire isomorphic class of rooted maps of the same genus that 
will be called a {\em rooted map}. 

A final remark: Walsh and Lehman (among many other authors) showed also that 
{\em counting rooted maps on the projective plane or on the sphere is 
topologically equivalent}. Although mathematically trivial, the consequence of 
this remark will be extensively used in this work while {\em displaying} the 
maps in figures: for the sake of simplicity, indeed, every rooted map on the 
sphere will be drawn as a rooted map on the plane.
\subsubsection{Euler-Poincar\'e invariant} One of the fundamental theorems of 
topology states that for any given map with $V$ vertices, $E$ edges and $F$ 
faces -- and embedded on a surface with $g$ holes -- there exists the following 
invariant:
\beq \label{poincare} V-E+F=2-2g=\chi(g),\ee
found as a generalization of the polyhedral formula; in Eq.(\ref{poincare}), 
 $\chi(g)$ is the {\em Euler characteristic} (or {\em Euler number}), 
sometimes also known as the Euler-Poincar\'e characteristic. The polyhedral formula is related to the number of polyhedron 
vertices $V$, faces $F$, and polyhedron edges $E$ of a simply connected (i.e., 
genus 0) polyhedron (or polygon).\footnote{It was discovered independently by Euler and 
Descartes in 1752 and it is also known as the Descartes-Euler polyhedral formula. 
The formula also holds for some, but not all, non-convex polyhedra and it has been generalized for 
$n$-dimensional polytopes by Schl\"afli in 1868.} Thus the genus of maps and rooted maps can be
derived from it. The only compact closed surfaces with Euler characteristic $0$ are the Klein 
bottle and torus (Dodson and Parker \cite{dodson}). 
We will often meet this concept in the following sections.
\subsection{Enumeration of rooted maps}
\subsubsection{Arqu\`es and B\'eraud approach}
Arqu\`es and B\'eraud, starting from Tutte's results, discovered a generating 
function for the rooted maps series \cite{Arques-rooted}. They proved that the 
generating series of rooted maps is the solution of  the following Riccati's 
differential equation:
$$ M(y,z)=y+zM(y,z)^2+zM(y,z)+2 z^2 \frac{\partial M(y,z)}{\partial z},$$
where $y$ and $z$ are respectively the number of vertices and edges. They have 
also shown that the generating series of rooted maps {\em with respect to the 
number of edges only} is solution of the differential equation:
$$ M(z)=1+zM(z)^2+zM(z)+2z^2\frac{\partial M(z)}{\partial z}.$$
Thanks to this relation, Arqu\`es and B\'eraud found the ``nice continued 
fraction form'' for the generating series $M(y,z)$ of rooted maps with respect 
to the number of vertices and edges:
$$ 
M(y,z)=\frac{y}{1-\frac{(y+1)z}{1-\frac{(y+2)z}{1-\frac{(y+3)z}{1-\ldots}}}},$$
with an important corollary: the generating series $M(y,z)$ of rooted maps with 
respect to vertices and edges is the solution of the following generalized 
Dyck's equation:
$$M(y,z) = y+zM(y,z)M(y+1,z).$$
By means of this corollary, the same authors found the explicit formula for 
the number of rooted maps with $n$ edges:
$$ 
M(n)=\frac{1}{2^{n+1}}\sum_{i=0}^{n}(-1)^i\underbrace{\sum_{k_1+\ldots+k_{i+1}
=n+1}}_{k_1,\ldots k_{i+1}\ge 0}\prod_{j=1}^{i+1} \frac{(2k_j)!}{k_j!}.$$
The anchor conditions which appear in the second summation show that we have to 
sum over all the possible partitions of the integer $n+1$ in $i+1$ integer 
numbers. In the following Table \ref{numeri}, the first terms of this formula 
are shown.
\begin{table}[ht]
\begin{minipage}[b]{0.46\textwidth}
\caption{The number of rooted maps $M(n)$ on an orientable surface as a 
function of the number of edges $n$ and regardless to genus, according to Arqu\`es 
and B\'eraud 
\cite{Arques-rooted},
 up to 4 edges.}
\begin{center}
{\begin{tabular}{cc} 
$n$ & $M(n)$  \\  \hline \\
0 & 1 \\
1 & 2 \\
2 & 10\\
3 & 74\\
4 & 706\\ 
\end{tabular}}
\end{center}
\label{numeri}
\vspace{3.7cm} 
\end{minipage}
\hspace{.5cm}
\begin{minipage}[b]{0.5\textwidth}
\small
\caption{The number of rooted maps with $e$ edges and $v$ vertices 
by genus $g$ up to 4 edges. From 
\cite{WI},
p. 215. The last (supplementary) 
column corresponds to Table \ref{numeri} and contains the first terms of the 
Arqu\`es-Walsh sequence: 1, 2, 10, 74, 706, \ldots}
\begin{center}
\begin{tabular}{r rrrrc} 
$e$ & $v$ & $g=0$ & $g=1$ & $g=2$ & $\sum$ {\em per edges}\\ \hline \\
0 & 1 & 1 & & &  \bf{1}\\
1 & 1 & 1 & & &  \bf{2}\\
  & 2 & 1 & & &   \\
2 & 1 & 2 & 1 & &  \bf{10}\\
  & 2 & 5 & & &  \\
  & 3 & 2 & & &  \\
3 & 1 & 5 & 10 & &  \bf{74}\\
  & 2 & 22 & 10 & &  \\
  & 3 & 22 & & &  \\
  & 4 & 5 & & & \\
4 & 1 &  14  &  70 & 21 &  \bf{706}\\
  & 2 &  93  & 167 & &  \\
  & 3 &  164 &  70 & &  \\
  & 4 &  93 & & &  \\
  & 5 & 14 & & &  \\
\end{tabular}
\end{center}
\label{numeridiwalsh1}
\end{minipage}
\end{table}

\subsubsection{Walsh and Lehman approach} Walsh and Lehman~\cite{WI} began their work 
 by studying the {\em combinatorial equivalent of maps}: this approach allowed 
 them to deal with the map enumeration 
from a combinatorial point of view (rather then a topological one). Then they 
gave a simple application of counting rooted maps regardless to genus. Later 
on, they generalized the Tutte's recursion formula \cite{Tutte1}
for higher genus for counting {\em slicings} and introduced the concept of {\em 
dicing} which actually is a ``contracted slicing'': in short, they considered 
the map obtained from slicings by contracting each band to a point. Thus a 
``dicing'' is a map whose vertices are distinguished by labeling each vertex 
with a different natural number ($d_i$). In order to summarize this important 
result, let us introduce the number of dicings of a genus $g$ surface whose 
vertices are of degree $d_1, \ldots d_v$: we define it as $C_g(d_1, \ldots 
d_v)$. The authors proved that the following recursion formula holds:
%
\bea
 C_g(d_1, \ldots d_v)&=& \sum_{i=1}^{v-1} d_i C_g (d_1, \ldots 
d_{i-1},d_i+d_v-2,d_{i+1}, \ldots d_{v-1})
+ \sum_{\underbrace{k+m=d_{v-2}}_{k\geq 0, m \geq 0}} C_{g-1}(d_1, \ldots 
d_{v-1}, k, m) \nonumber \\
 &+& \sum_{\underbrace{D_1 \cup D_2=d_1+ \ldots d_{v-1}}_{D_1 \cap D_2=\phi}} 
\sum _{h+f=g}
\sum_{\underbrace{k+m=d_{v-2}}_{k\geq 0, m \geq 0}} C_h(D_1,k) C_f(D_2,m). 
\label{Recu}
\eea
%

%
This formula reduces to the Tutte's recursion formula~\cite{Tutte1}
when $g=0$. For further details, such as the explicit combinatorial meaning of 
dicings or the proof of uniqueness of the solution of the Walsh and Lehman's 
recursion formula, see \cite{WI}.  By means of (\ref{Recu}), they computed the 
number of dicings given the degree of each vertex and the genus. In particular, 
Walsh and Lehman extracted an explicit formula for maps with one face. Finally, 
they obtained a relation between the number of dicings and the number of rooted 
maps. 
This relation allows to calculate the number of rooted maps with respect 
to the genus, the number of vertices and the number of edges. In other words, 
they obtained the following relation: if we denote with $C_g(d_1, \ldots 
d_v)$\footnote{The problem actually is to find an explicit expression for the 
term $C_g(d_1, d_2, \ldots d_v)$. Walsh and Lehman for instance found an 
explicit form for the rooted maps with $one$ face, where 
$d_1+\ldots+d_v=4g+2v-2$, which leads to the very interesting formula:
$$ M(n,v,g)_{\textrm{one face}}=\frac{(2v+4g-2)!}{2^{2g}v!(v+2g-1)!}
\underbrace{\sum_{i_1+\ldots+i_v=g}}_{i_1,\ldots i_v\ge 0}\prod_{j=1}^v 
\frac{1}{1+2i_j}.$$
This is the number of rooted maps of genus $g$ with one face, $v$ vertices and 
$v+2g-1$ edges and, by duality between vertices and faces, it is also the number 
of rooted maps of genus g with one vertex, $v$ faces and $v+2g-1$ edges. Compare 
this expression with the $M(n)$ of the next section.} the number of dicings and 
an explicit expression is
known for it, the number of rooted maps of genus $g$, with $v$ vertices and $e$ 
edges is:
$$ \frac{2e}{v!} \sum_{d_1+ d_2 + \ldots d_v=2e} \frac{C_g(d_1, d_2, \ldots 
d_v)}{\prod_{i=1}^{v}d_i}.$$
Summing over all the descending sequences $(d_1, \ldots d_v)$ which add to $2e$ 
gives the number of rooted maps of genus $g$ with $e$ edges  and $v$ vertices. 
In this way they were able to fill the Table I 
of Ref.~\cite{WI}. An 
extract of this result is shown in Table \ref{numeridiwalsh1}. Interestingly, if 
we sum the number of maps with $n$-edges (i.e. independently of the genus $g$), 
we obtain the sequence 1, 2, 10, 74, 706, \ldots i.e. the number of rooted maps 
as a function of the number of edges and regardless to genus, as found by 
Arqu\`es and B\'eraud \cite{Arques-rooted} (see Table \ref{numeri}). We will 
refer to this sequence as to the {\bf Arqu\`es-Walsh sequence}.
\section{Feynman diagrams and rooted maps}\label{sec3}
As anticipated in the Introduction, the main purpose of the present work lies in 
the connection between counting rooted maps and enumerating Feynman diagrams in the 
perturbative series which expresses the single particle Green's function (or propagator) 
within the non-relativistic many-body theory of a system of fermions (particles with 
half-integer spin). 

Without any pretense of completeness, we recall here only the starting point, key 
formulas and definitions, which lead to the perturbative expression one can represent 
in terms of Feynman diagrams. The reader is addressed, e.g., to the Fetter and Walecka 
texbook~\cite{Wale} for a comprehensive derivation of the pertinent formulas.

The physical system is described by an Hamiltonian operator 
$$\hat H = \hat H_0 + \hat H_1$$
where $\hat H_0$ corresponds to a non-interacting (solvable) system, while $\hat H_1$
contains the (two-body) interaction between the constituents of the system.

The single-particle Green's function is then defined by
\begin{equation}
 i\mathcal{G}_{\alpha \beta}(x,y)=\frac{<\Psi_0|T[\hat\Psi_{H_\alpha}(x)\hat\Psi^\dagger_{H_\beta}(y)]|\Psi_0>}
 {<\Psi_0|\Psi_0>}
 \label{Green-def}
\end{equation}
where $|\Psi_0>$ is the exact ground state of the system, $\hat\Psi_{H_\alpha}$ ($\hat\Psi^\dagger_{H_\beta}$)
are field operators in Heisenberg representation, destroying (creating) a particle in a specific space-time
point with the appropriate spin projection. $T$ is the time-ordered product, forcing the operator with the latest 
time to be placed on the left. The quantity in Eq.(\ref{Green-def}) also represents the propagation of the 
interacting particle from point $x\equiv (\mathbf{x},t_x)$ to  $y\equiv (\mathbf{y},t_y)$ (or vice versa).

Perturbation theory leads then, with the help of a few celebrated quantum fields theorems, to the expression:
\begin{eqnarray}  \label{newGreen}
 &&i\mathcal{G}_{\alpha \beta}(x,y)= \sum_{m=0}^\infty \left(-\frac{i}{\hbar}\right)^m \frac{1}{m!}
 \int_{-\infty}^{+\infty}dt_1 \cdots \int_{-\infty}^{+\infty}dt_m
 \\ \nonumber
 && \times<\Phi_0|T[\hat H_1(t_1)\cdots \hat H_1(t_m) \hat\Psi_\alpha(x) \hat\Psi^\dagger_\beta(y)]|\Phi_0>_{\mathrm{connected}}
\end{eqnarray}
where $m$ is the order of the perturbative term and all operators are in Interaction representation, $|\Phi_0>$ being
now the ground state of the non-interacting system. Notably it can be shown that the series extends only to the 
connected terms, namely those terms where the interaction Hamiltonians are connected to the ``fixed'' external points 
$x,y$, or equivalently to the fermion propagator running from $y$ to $x$. The terms of Eq.(\ref{newGreen}) and their enumeration
can be put into a one to one correspondence with Feynman diagrams, according to the rules explained in the Appendix.

A detailed enumeration of Feynman diagrams according to their physical 
properties (and as a function of the perturbative order) is presented in the 
work of Cvitanovi\'c, Lautrup and Pearson~\cite{C}. 
In particular, Table I 
contains the reckoning of several subtypes of QED diagrams. The first two columns show the effect of the 
Furry's theorem which cancels a large number of diagrams in QED. Conversely, in 
many-body theory, all the diagrams which appear in the first column (``Exact 
electron propagators without Furry's theorem'') can contribute to the total 
amplitude\footnote{In this Table, Cvitanovi\'c, Lautrup and Pearson use a 
different notation: they refer to the perturbative order of a Feynman diagram as 
to the number of interaction vertices: this is indeed customary in QED, where 
the fundamental interaction Lagrangian refers to the electron-photon coupling. 
In this framework the exchange of a photon between two electrons is viewed as a 
second order term.}. The reason why in many-body theory the Furry's theorem is 
not active lies in the different nature of the {\em vacuum} in the two 
approaches: the Dirac sea with an infinite number of negative energy states in 
QED and the (unperturbed) ground state of $N$ particles in many-body theory 
(Fermi sea). Clearly vacuum loops diverge in QED and are removed by standard 
renormalization procedures. Since these divergences do not occur in many-body 
theory, the number of Feynman diagrams (as a function of the perturbative order 
for the exact electron propagator) without taking into account Furry's theorem 
is
$$1, \, 2, \, 10, \, 74, \, 706\ldots$$
 Remarkably, {\bf it corresponds to the Arqu\`es-Walsh sequence}, i.e. to the 
number of rooted maps regardless to genus and vertices as a function of the 
number of edges. 
The numerical correspondence is quite striking and it is suggestive that 
between this two objects, the physical ones (Feynman diagrams) on the one side and 
the mathematical ones (rooted maps) on the other side, there exists a topological 
connection. Our aim is then to explore this topological equivalence. This goes 
beyond a simple {\em counting} of these objects (which corresponds to compare the Arqu\`es 
formulas on pages 6-10 of \cite{Arques-rooted} with the ones of Cvitanovi\'c 
\cite{C} on page 1943). The purpose is rather to discover {\em how} a Feynman 
diagram is related to a rooted map. It is in fact the {\em topology} of these 
objects which entails their {\em physical} content (while an exact match between 
these formulas is rather of a mathematical interest).
\subsection{Connection between diagrams and maps}
\begin{figure}
\begin{minipage}[b]{0.37\textwidth}
\centering
\includegraphics[width=0.95\textwidth]{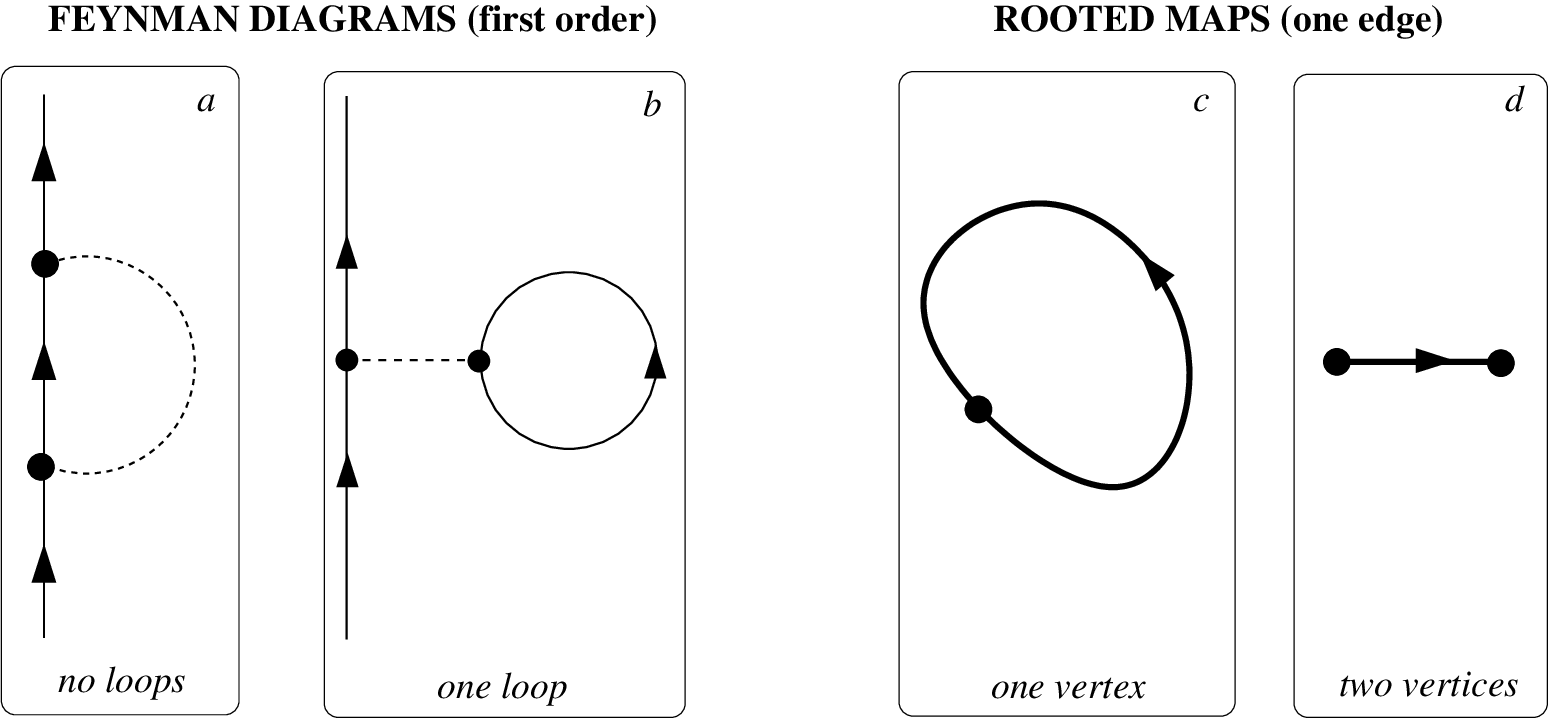}
\caption{The easiest (non-trivial) Feynman diagrams and (non-trivial) rooted 
maps as they appear in Refs.~\cite{Wale} and \cite{Tutte2}, respectively. Interaction 
lines will be represented with dashed lines, propagation lines with solid ones. } 
\label{1to1.0}
\end{minipage}
\hspace{0.1cm}
\begin{minipage}[b]{0.57\textwidth}
\centering
\includegraphics[width=0.99\textwidth]{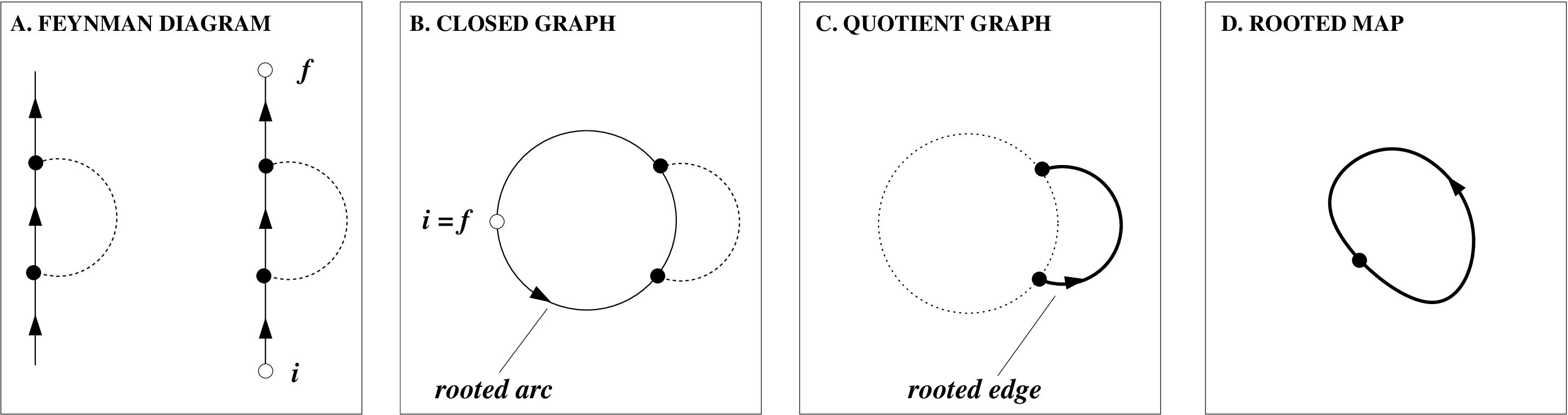}\\
\vspace{0.5cm}\includegraphics[width=0.99\textwidth]{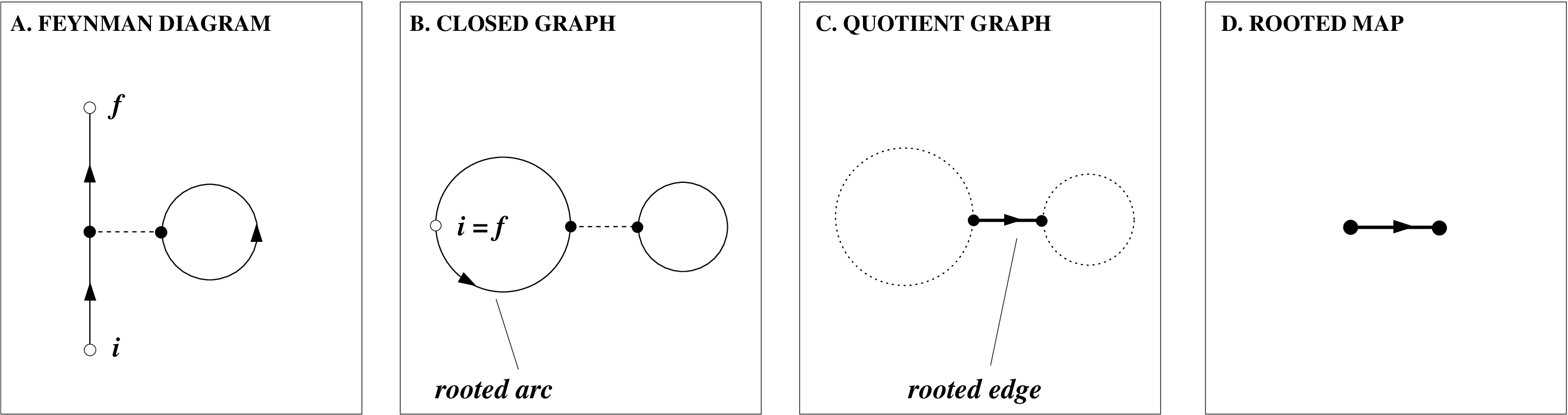}
\caption{(Top) Association of the first-order Feynman diagram $a$ to the rooted 
map $c$ of Fig.\ref{1to1.0} by means of the {\em quotient procedure}. (Bottom) 
Association of the first-order Feynman diagram $b$ to the rooted map $d$ of 
Fig.\ref{1to1.0}. The ``initial'' ($i$) and ``final'' ($f$) points are shown.} 
\label{1to1.1}
\vspace{.3cm}
\end{minipage}
\end{figure}

On the {\em physical} side, the starting point is the book Ref.~\cite{Wale},
where the Feynman diagrams (at first and second order) are explicitly 
drawn\footnote{The work of Cvitanovi\'c \cite{C} 
will help us to enumerate and distinguish the physics properties of Feynman 
diagrams at any order.}. 
On the {\em mathematical} side, the starting point is the work of Tutte~\cite{Tutte2}, 
where the (planar) rooted maps (up to 2 
edges) are also explicitly drawn\footnote{ The work of Walsh and Lehman 
\cite{WI} (summarized in Table \ref{numeridiwalsh1}) will help us with the 
detailed enumeration of the 
rooted maps according to their number of vertices, edges and the genus of the 
embedding surface.}. Fig.\ref{1to1.0} shows the easiest (non-trivial) examples 
of Feynman diagrams (left) at first order of perturbation: diagram $a$ (also 
known as ``shell'') contains no loops while diagram $b$ (also known as 
``tadpole'') contains one loop. Fig.\ref{1to1.0} (right) shows also the first 2 
non-trivial rooted maps with one edge and one vertex (map $c$) and two vertices 
(map $d$) as they appear in Ref.~\cite{Tutte2}. We will use the two diagrams and the 
two maps to illustrate a graphical, step-by-step association (which will be indicated as the {\em quotient 
procedure}) -- for the first order diagrams (or for the rooted map with one 
edge) -- between these two kinds of objects.

\subsubsection{Quotient procedure (first order)} The following steps illustrate an 
intuitive way to associate a rooted map to a given Feynman diagram (at first 
order). They do not represent a rigorous proof of the association, but they will 
help us to investigate the topological equivalence between diagrams and maps. As 
a preliminary remark, we recall that for each Feynman diagram, an ``initial'' 
point and a ``final'' point can always be defined without any ambiguity: these 
points represent the initial and final positions in spacetime between which we evaluate the Green's 
function\footnote{These points are easily distinguished in the diagram, since 
they are the only vertices of a propagation line that are not connected to an 
interaction line.}.

\begin{itemize}
\item Starting from a Feynman diagram (step {\bf A} of Fig.\ref{1to1.1}), we 
connect its initial ($i$) and final ($f$) points, obtaining a new object: the 
{\bf closed graph} (step {\bf B}). The obtained (oriented) arc will be defined 
as the {\em rooted arc}\footnote{The rooted arc can be obtained either 
connecting $i$ and $f$ lefthandwise or righthandwise (so that the rooted arc 
results oriented counterclockwise or clockwise, respectively). This choice will 
not affect the results, provided that we apply the same rule for all the 
diagrams. Here we will use the first one.}. All the propagation-line arrows will 
be left out, except for the one on the rooted arc.
\item Starting from the point $i = f$ in the closed graph, we travel the rooted 
arc (e.g. counterclockwise) till the first interaction line and we shift the 
rooted arc arrow on it. Then, all the interaction lines will be drawn as solid 
lines and all the propagation lines will be drawn as dotted lines, obtaining the 
{\bf quotient graph} (step {\bf C}). The (unique) solid line carrying the arrow 
will be referred as to the {\em rooted edge}.
\item All the dotted lines in the quotient graph are collapsed to a unique 
vertex each. The result is a rooted map (step {\bf D}).
\end{itemize}
\begin{figure}
\begin{minipage}[b]{0.47\textwidth}
\centering
\includegraphics[width=0.95\textwidth]{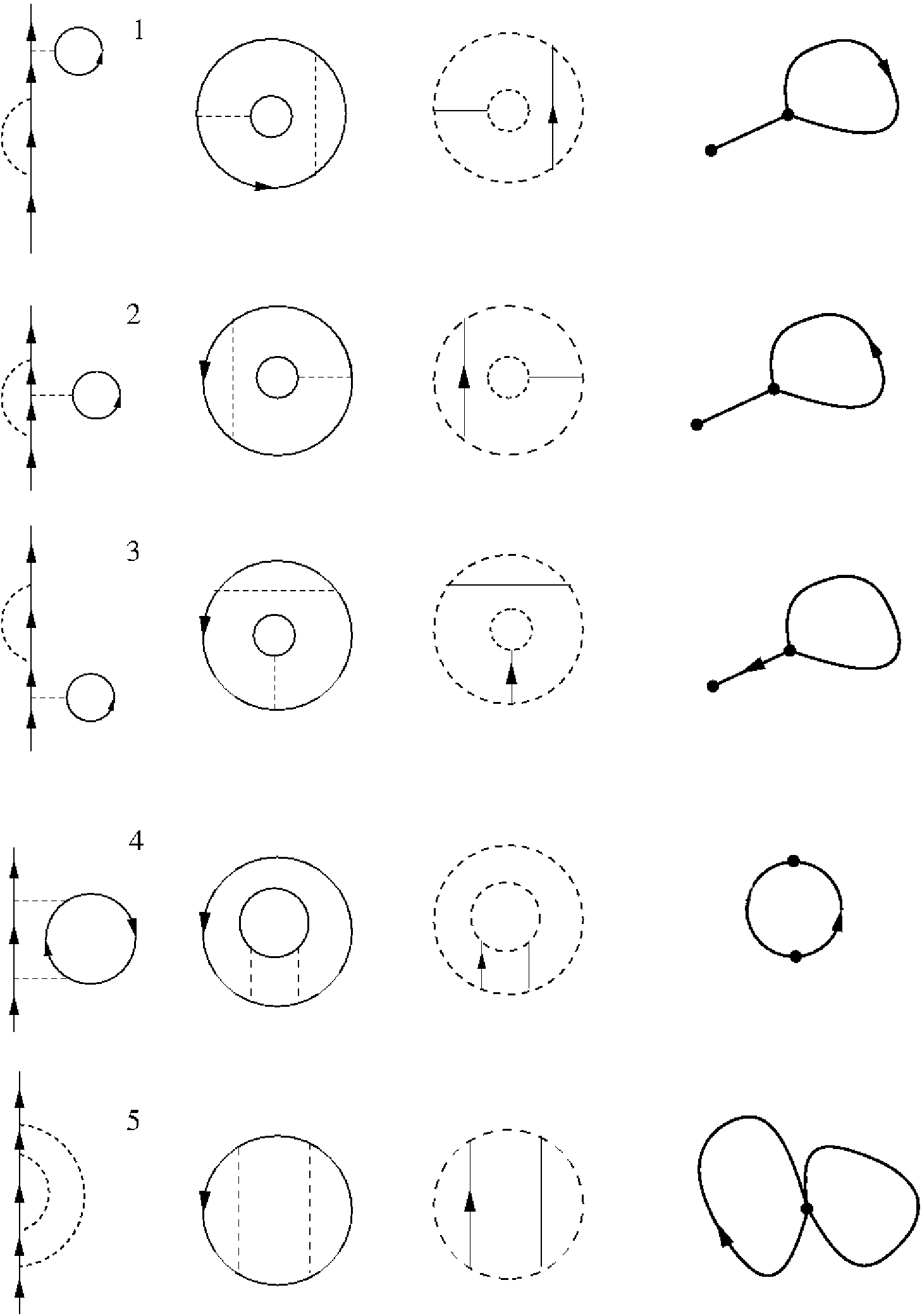}
\caption{Association of the second-order Feynman diagrams (as they appear in 
\cite{Wale}) with the two-edges rooted maps (on the sphere) as they appear in 
\cite{Tutte2}
by means of the {\em quotient procedure}. In the first column, the 
Feynman diagrams at second perturbative order (i.e. with 2 interaction lines) 
are shown. In the second and third columns the related closed graphs (with their 
rooted arc) and quotient graphs (with their rooted edge) are shown. The last 
column contains the resulting rooted maps. The list continues in Fig.\ref{IIorder2}.}
\label{IIorder1}
\end{minipage}
\hspace{0.1cm}
\begin{minipage}[b]{0.47\textwidth}
\centering
\includegraphics[width=0.95\textwidth]{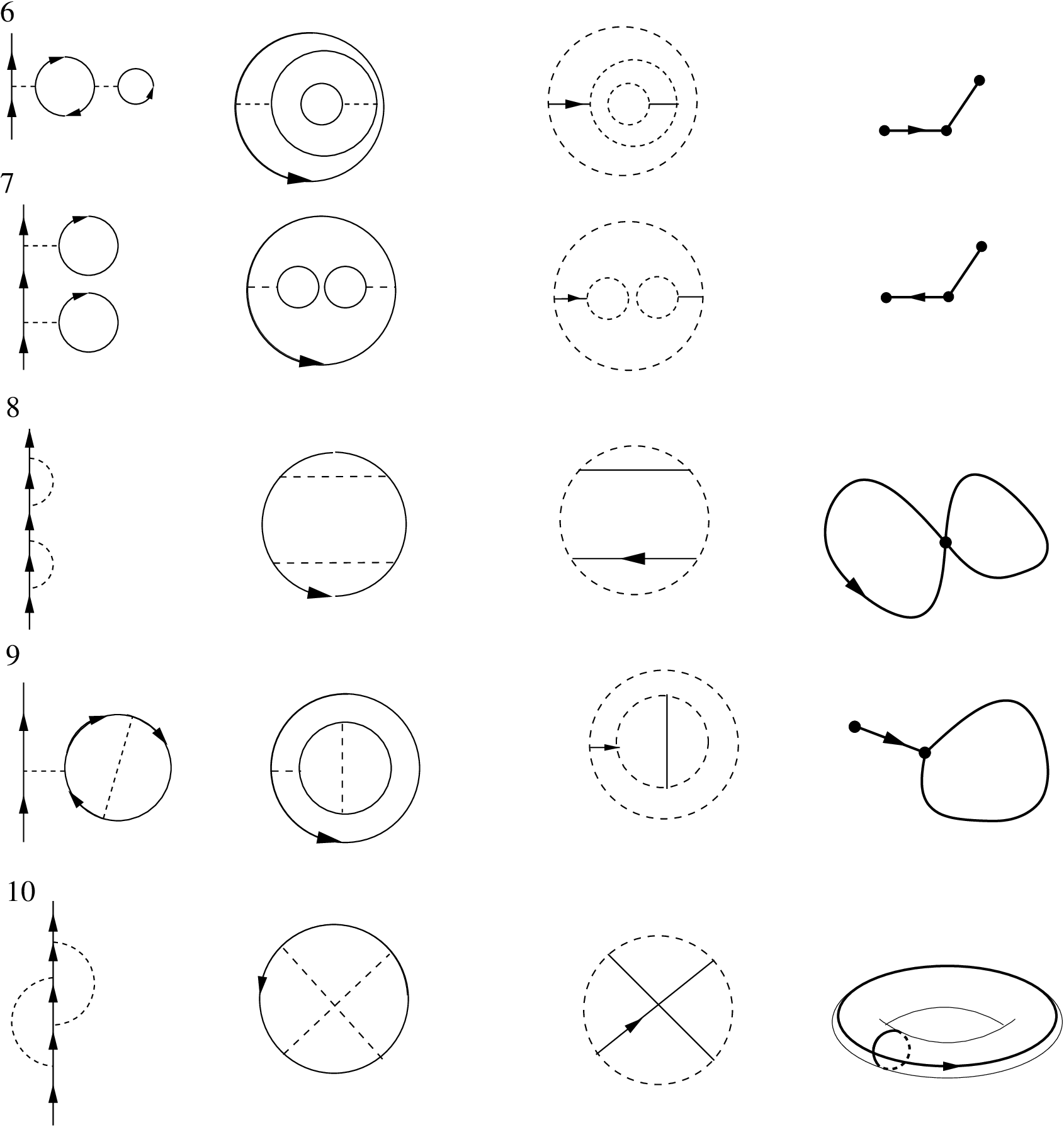}
\caption{Continues from Fig.\ref{IIorder1}. Notably, the rooted map related to 
the Feynman diagram n.10 is embedded on a torus.}
\label{IIorder2}
\vspace{3cm}
\end{minipage}
\end{figure}

\subsubsection{Quotient procedure (second and higher order)} 
\begin{figure}
\begin{minipage}[b]{0.47\textwidth}
\centering
\includegraphics[width=0.95\textwidth]{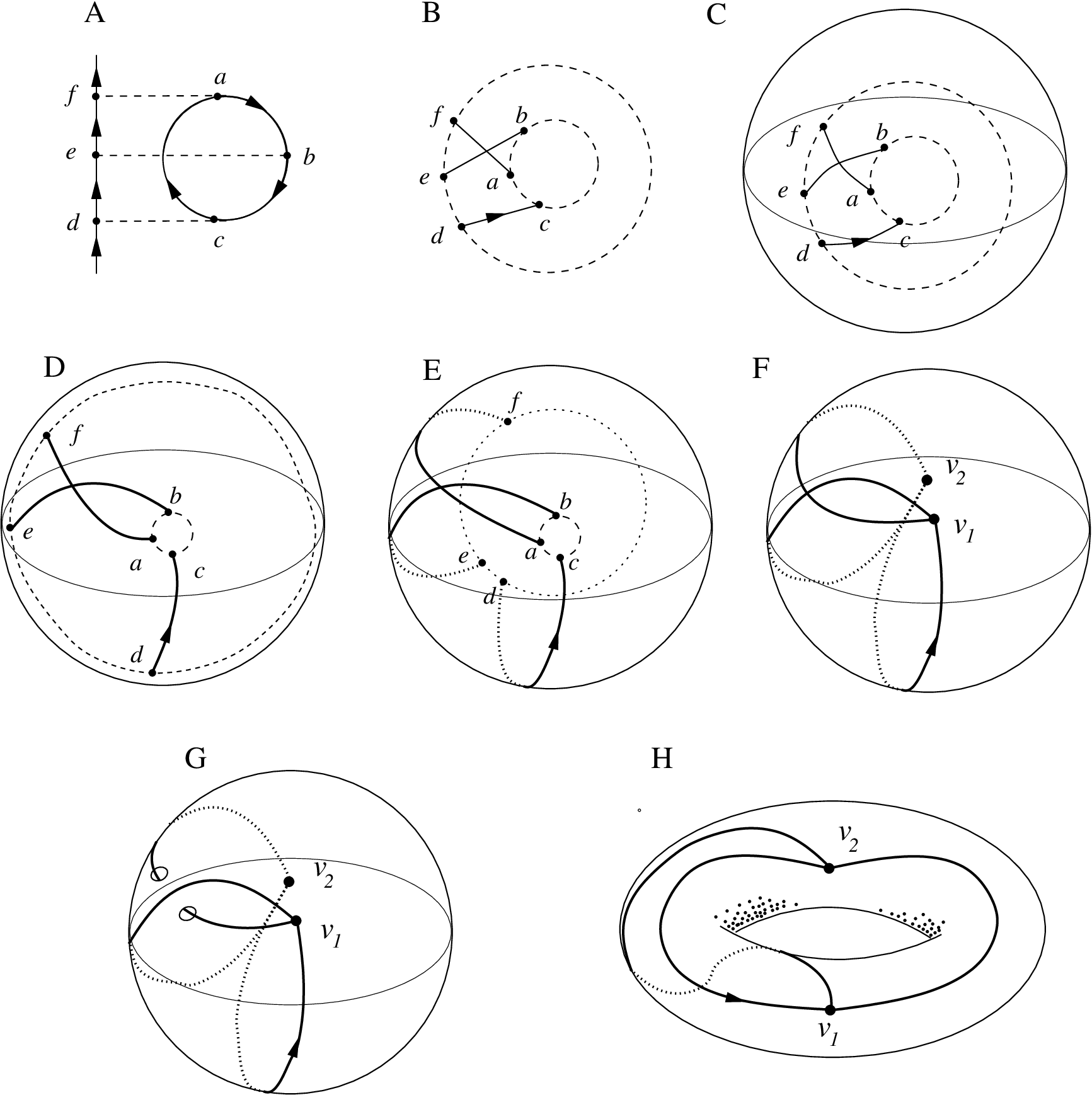}
\caption{The quotient procedure applied to a third-order Feynman 
diagram. Once we have embedded its quotient graph on an orientable surface (for 
example a sphere, frame C), we collapse the dashed lines into vertices (here 
$v_1$ and $v_2$), while maintaining the relative positions of the 
interaction-line extremes $a, b, c \ldots$ on their dashed lines (see pictures 
D-E-F). In order to remove the crossings between the resulting edges, we add a 
hole into the surface. }\label{nice1}
\end{minipage}
\hspace{0.1cm}
\begin{minipage}[b]{0.47\textwidth}
\centering
\includegraphics[width=0.95\textwidth]{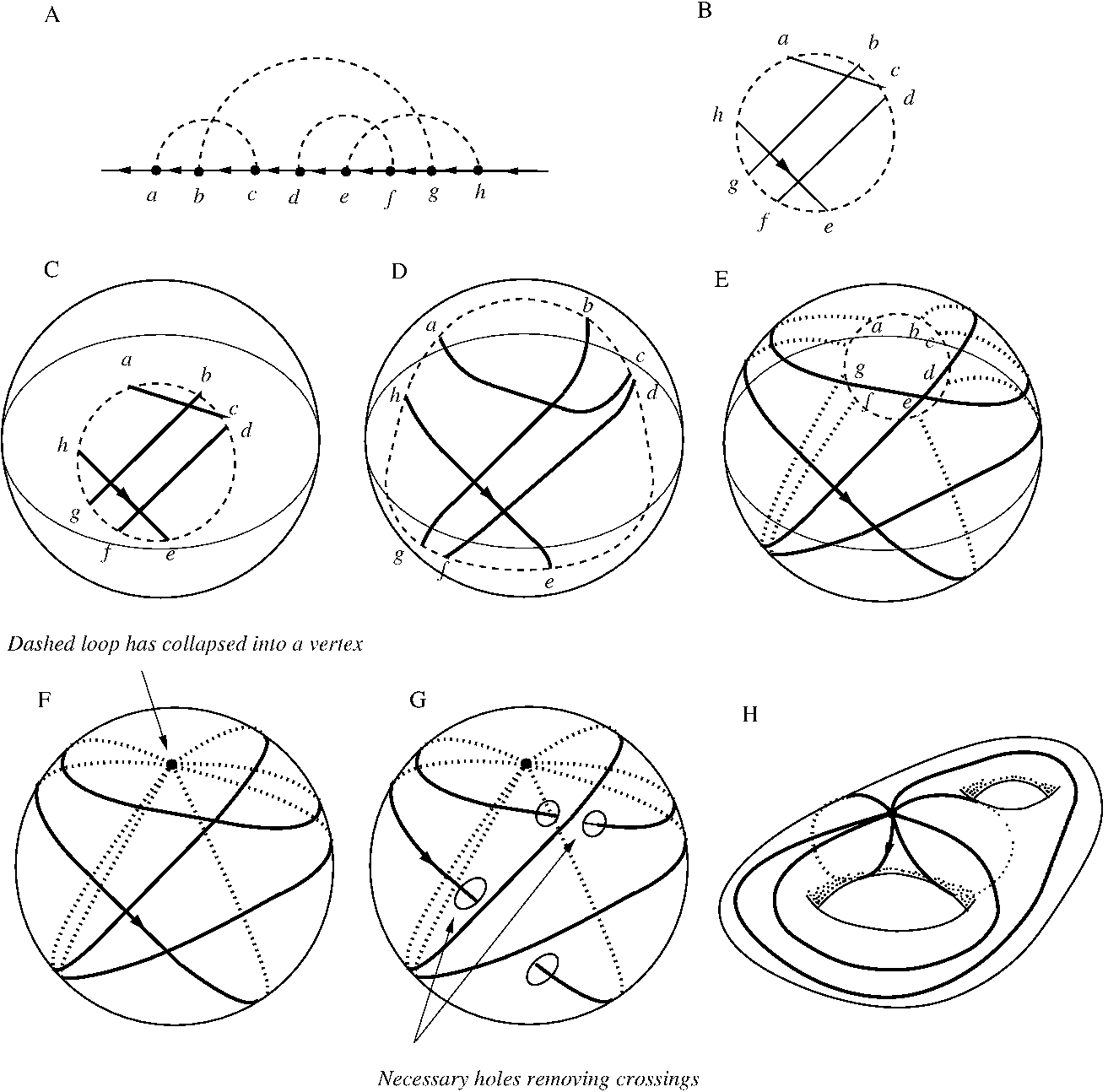}
\caption{The quotient procedure applied to a fourth-order Feynman 
diagram. In order to remove the crossings in the quotient graph, one hole is not 
enough. The associated rooted map results actually embedded on the 
$cow$.}\label{nice2}
\vspace{1.5cm}
\end{minipage}
\end{figure}

If we try to apply the previous 
steps to the second order diagrams in Ref.~\cite{Wale} (see Fig.\ref{IIorder1} and 
\ref{IIorder2}), we have no difficulties till the diagram number 10. The 
peculiarity of such a diagram can be seen under two perspectives. First, its 
quotient graph contains a {\bf crossing} (such a characteristic never occurred 
in all the other graph). Second, we already know that there should be $nine$ 
(planar) rooted maps with two edges (these are the maps drawn by Tutte): but we 
also know (see again Table \ref{numeridiwalsh1}) 
that one {\em and only one} rooted map with two edges should be embedded {\bf on 
the torus}. This means that another step  must be added to the quotient 
procedure: the quotient graph should be embedded on an orientable surface {\em 
with the minimum number of holes as it is needed to remove all the crossings} (step {\bf E}). 
We show in Figs~\ref{nice1} and \ref{nice2} some intriguing examples of order higher than two.

\subsubsection{Quotient procedure (full third order)}
\begin{figure}
\begin{minipage}[b]{0.47\textwidth}
\centering
\includegraphics[width=0.95\textwidth]{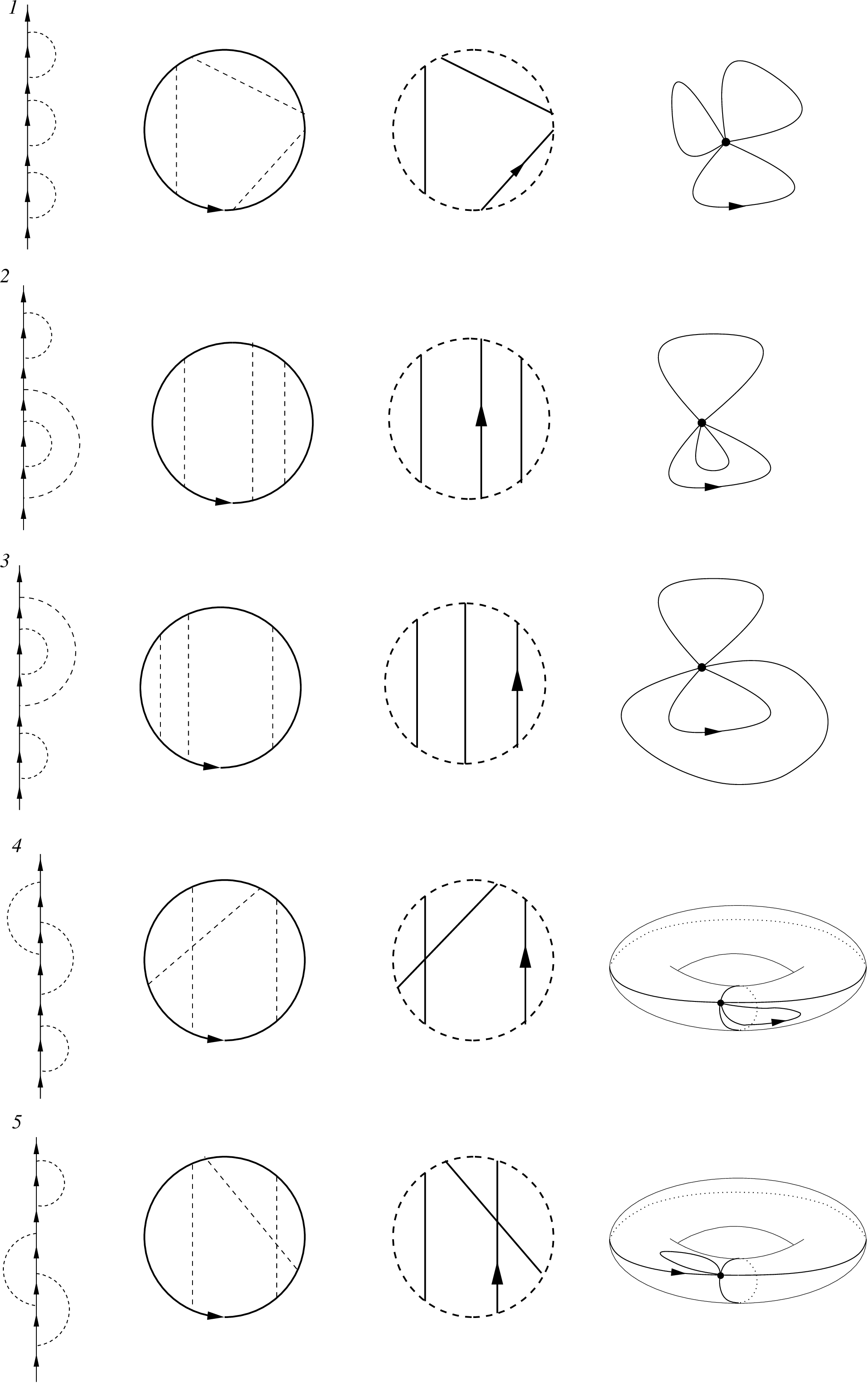}
\caption{Closed and quotient graphs of ``shells`` Feynman diagrams  and their 
related rooted maps, obtained by means of the quotient procedure. Feynman 
diagrams which are not embedded on a sphere are related to a map on the 
torus.}\label{quotshell1}
\end{minipage}
\hspace{.1cm}
\begin{minipage}[b]{0.47\textwidth}
\centering
\includegraphics[width=0.95\textwidth]{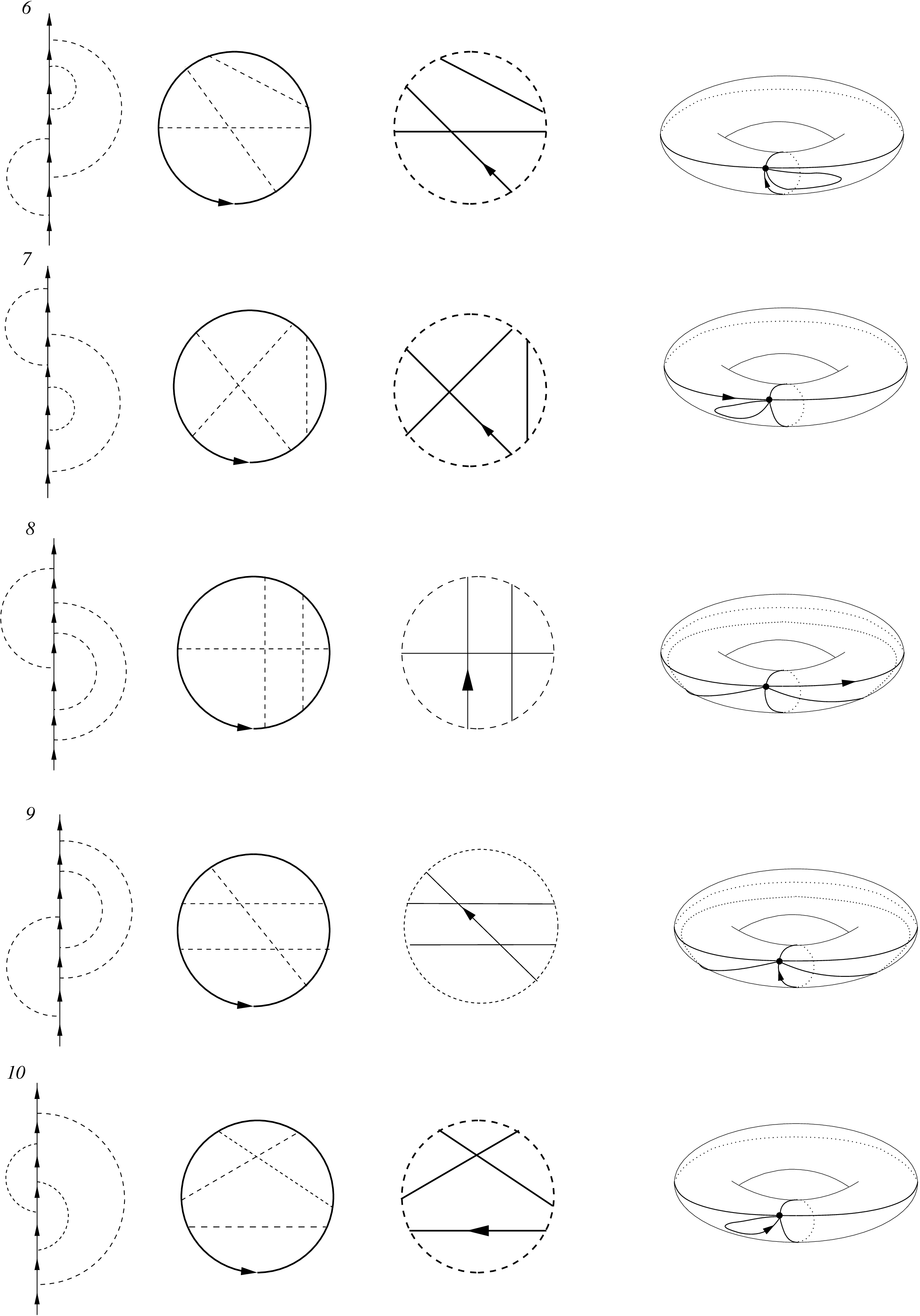}
\caption{Closed and quotient graphs of ``shells`` Feynman diagrams  and their 
related rooted maps ({\em continued}).}\label{quotshell2}
\vspace{0.7cm}
\end{minipage}
\end{figure}

\begin{figure}
\begin{minipage}[b]{0.47\textwidth}
\centering
\includegraphics[width=0.8\textwidth]{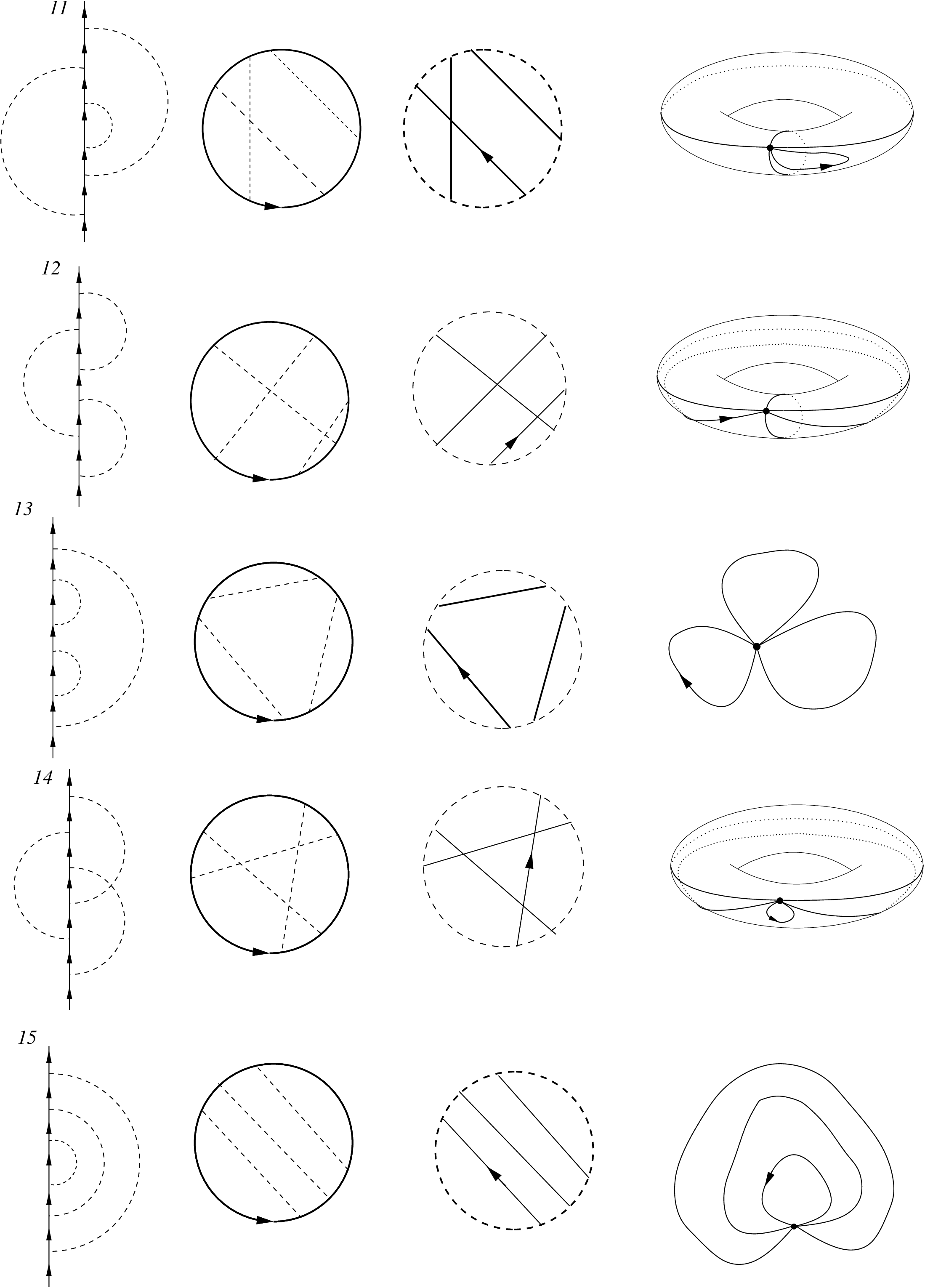}
\caption{Closed and quotient graphs of ``shells`` Feynman diagrams  and their 
related rooted maps ({\em continued}).}\label{quotshell3}
\end{minipage}
\hspace{0.1cm}
\begin{minipage}[b]{0.47\textwidth}
\centering
\includegraphics[width=0.98\textwidth]{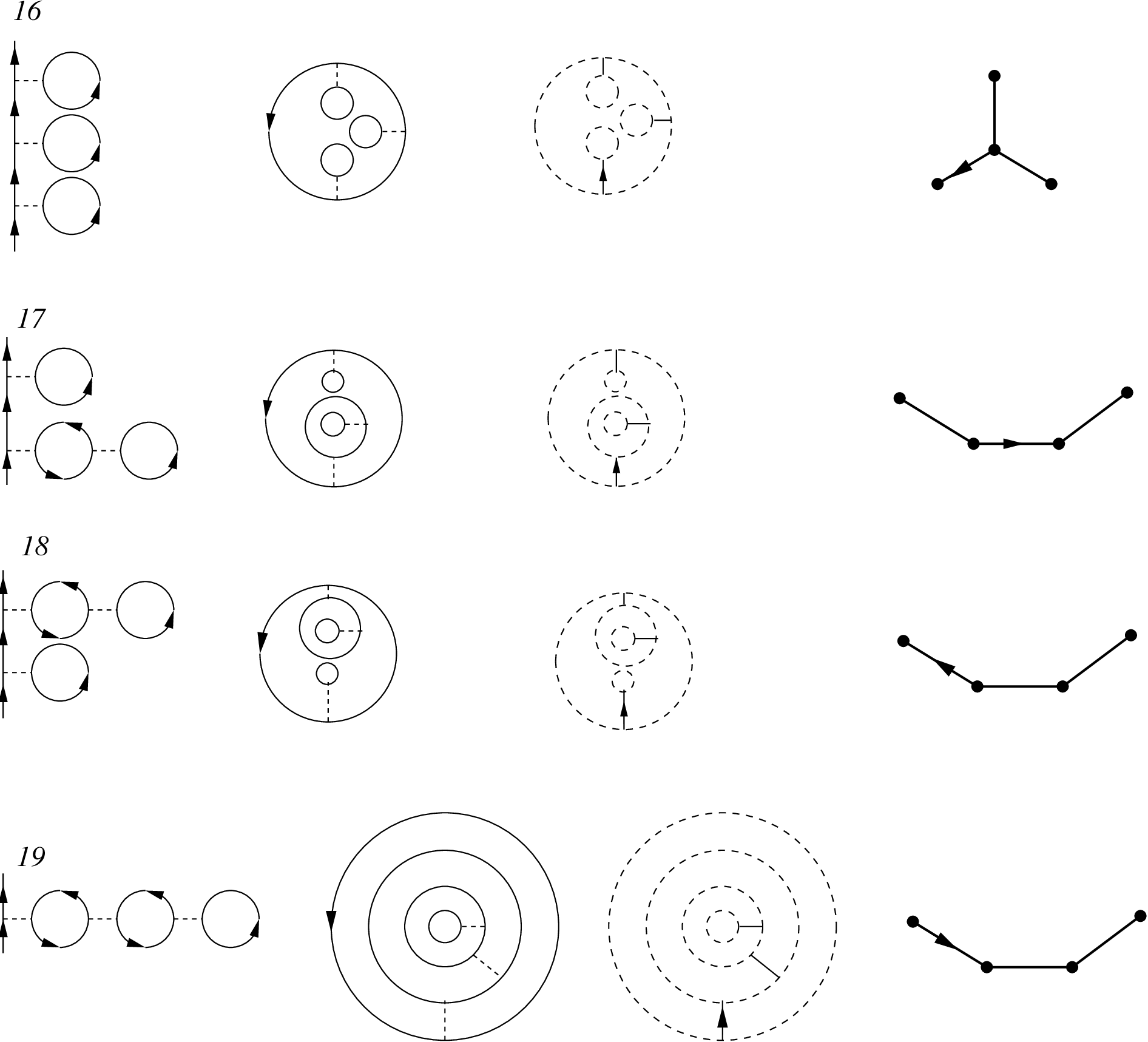}
\caption{Closed and quotient graphs of other third order diagrams and their 
related rooted maps.}\label{quot1}
\end{minipage}
\end{figure}

Let us consider the third order diagrams derived in the Appendix\footnote{In the Appendix, 
a simple method to build Feynman diagrams at any order is also presented. It was 
needed to recover the exact shape of each of the 74 diagrams at third order. 
Actually, a publication where these 74 Feynman diagrams are explicitly drawn is not available, 
at variance with the second order diagrams, which can be found, e.g., in Ref.~\cite{Wale}).}. 
We have first applied the quotient procedure to all the shell diagrams shown in Fig.
\ref{shells}. We can immediately observe that there are 10 diagrams which 
contain unremovable crossings between the interaction lines (we are talking 
about number 2, 3, 6, 7, 8, 9, 10, 11, 12, 14) and 5 diagrams without crossings 
(number 1, 4, 5, 13 and 15). Thanks to Table \ref{numeridiwalsh1}, we could 
predict that there could be only {\em five} diagrams on the sphere. If we apply 
the quotient procedure to the diagrams of Figs~\ref{quotshell1}, \ref{quotshell2}  and \ref{quotshell3}, 
we find indeed five and only five rooted maps on the sphere (diagrams number 1, 2, 3, 13 and 15 in the 
above mentioned figures).
\begin{figure}
\begin{minipage}[b]{0.47\textwidth}
\centering
\includegraphics[width=0.7\textwidth]{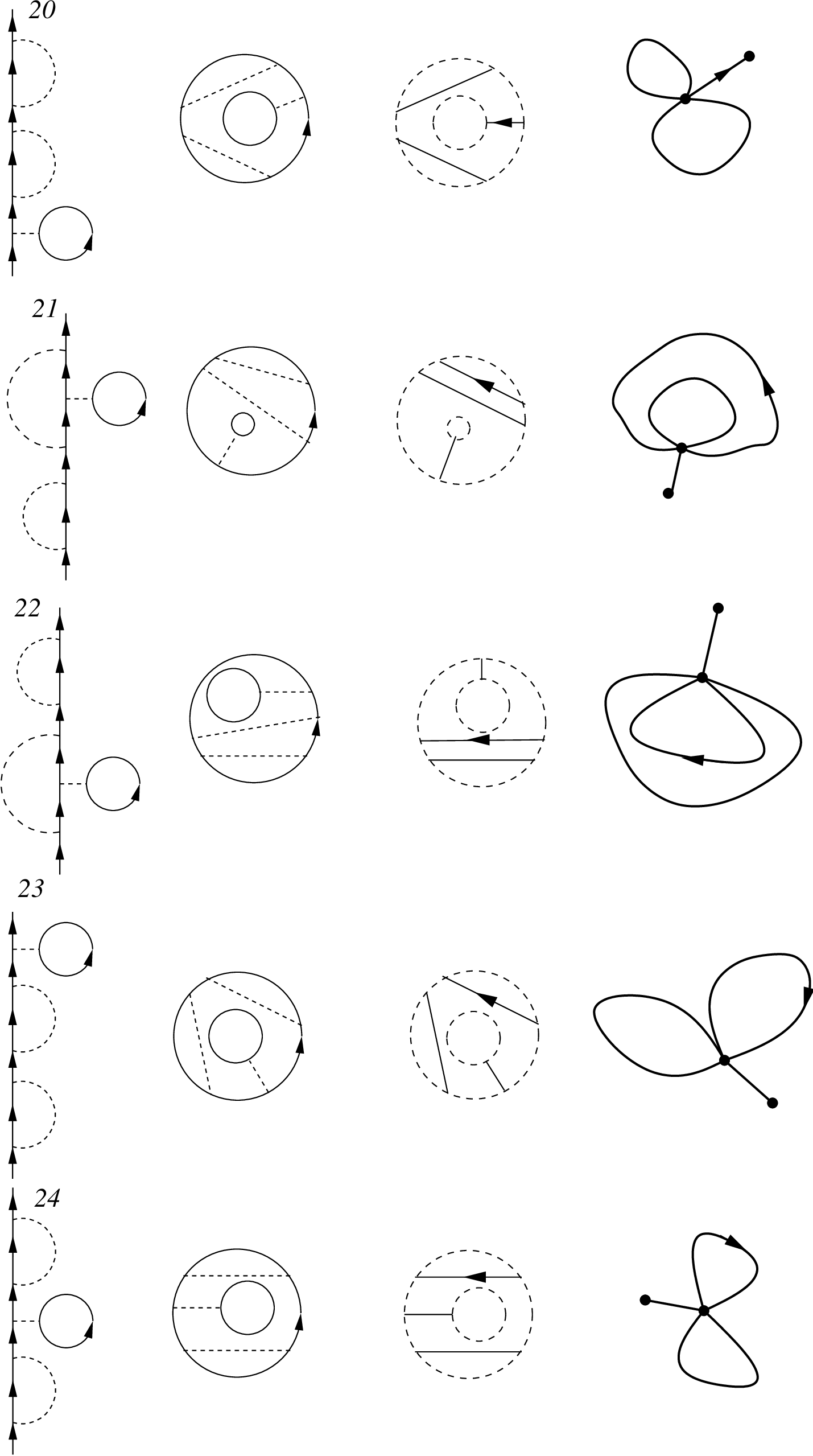}
\caption{Closed and quotient graphs of other third order diagrams and their 
related rooted maps.}\label{quot2}
\end{minipage}
\hspace{0.5cm}
\begin{minipage}[b]{0.47\textwidth}
\centering
\includegraphics[width=0.8\textwidth]{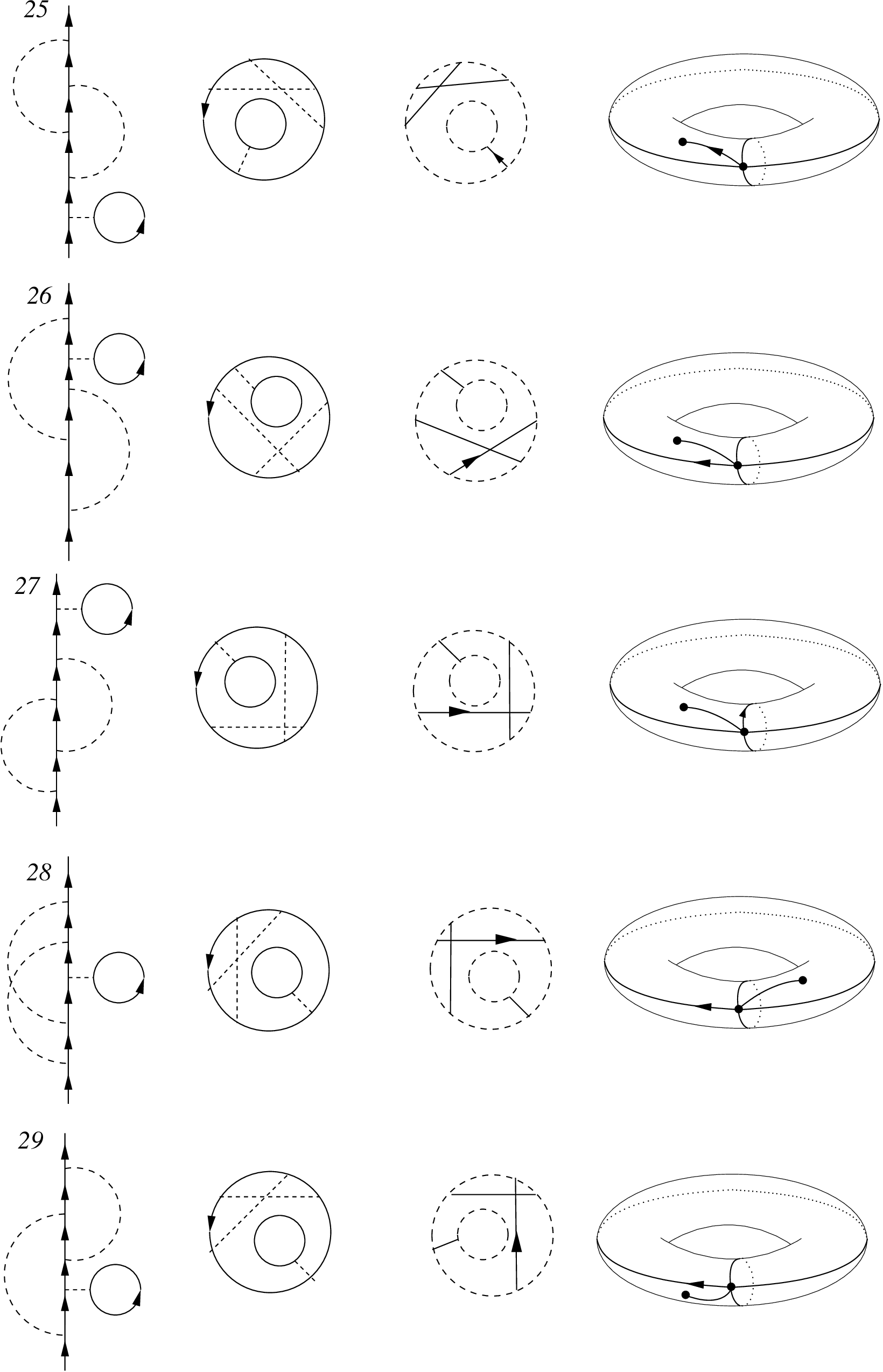}
\caption{Closed and quotient graphs of other third order diagrams and their 
related rooted maps.}\label{quot3}
\end{minipage}
\end{figure}
\begin{figure}
\begin{minipage}[b]{0.47\textwidth}
\centering
\includegraphics[width=0.7\textwidth]{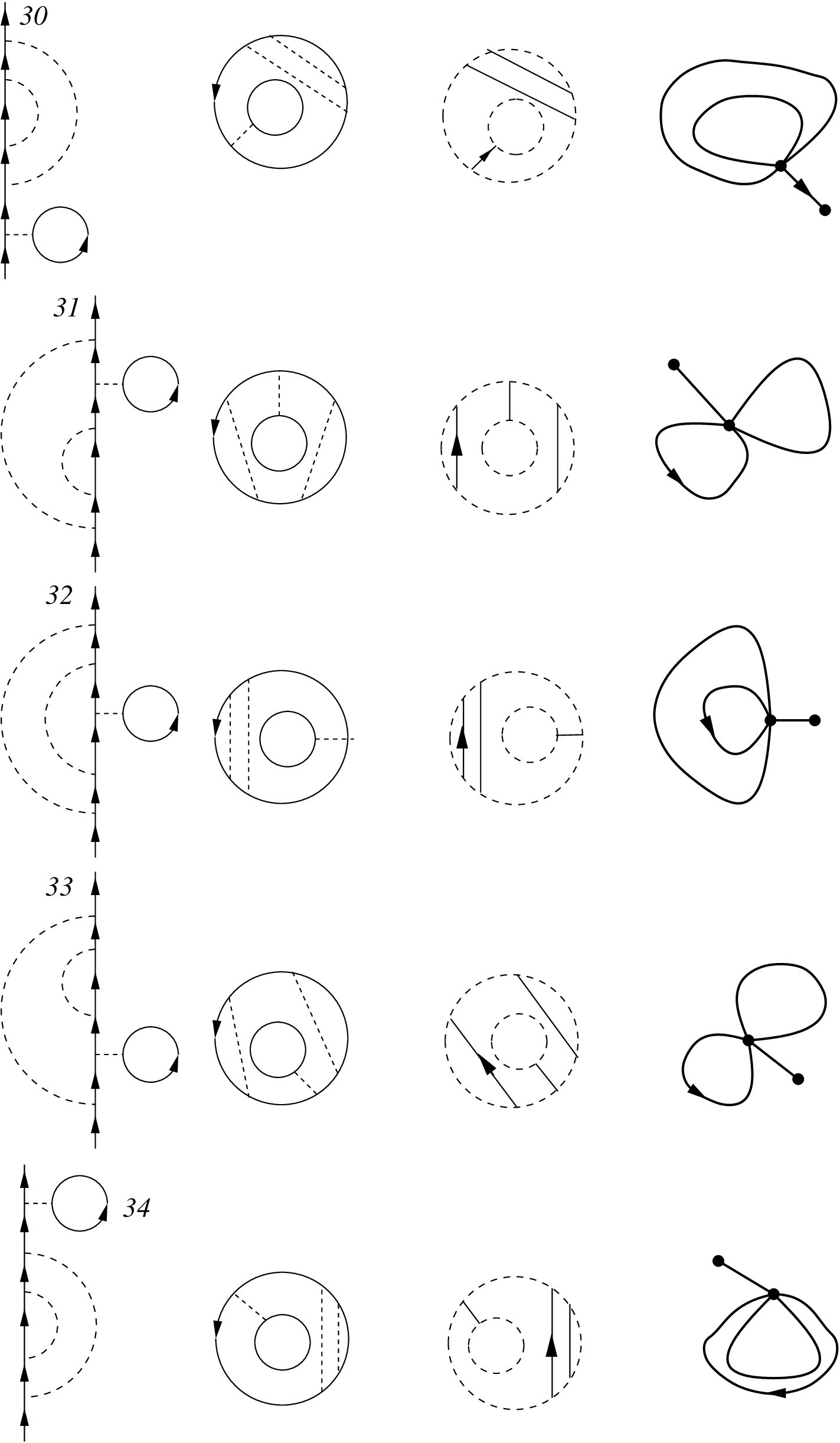}
\caption{Closed and quotient graphs of other third order diagrams and their 
related rooted maps.}\label{quot4}
\end{minipage}
\hspace{0.5cm}
\begin{minipage}[b]{0.47\textwidth}
\centering
\includegraphics[width=0.8\textwidth]{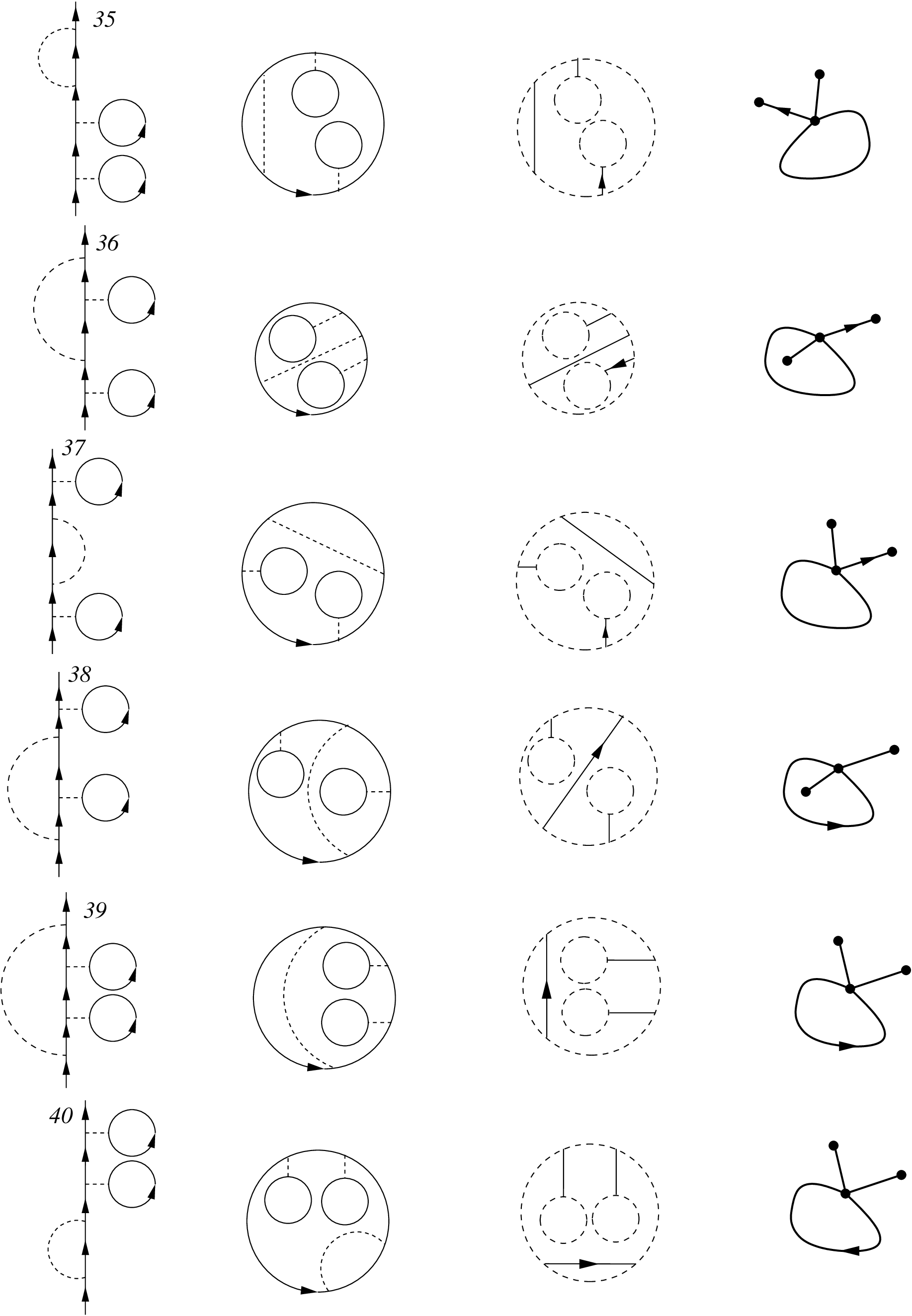}
\caption{Closed and quotient graphs of other third order diagrams and their 
related rooted maps.}\label{quot5}
\end{minipage}
\end{figure}
\begin{figure}
\begin{minipage}[b]{0.47\textwidth}
\centering
\includegraphics[width=0.7\textwidth]{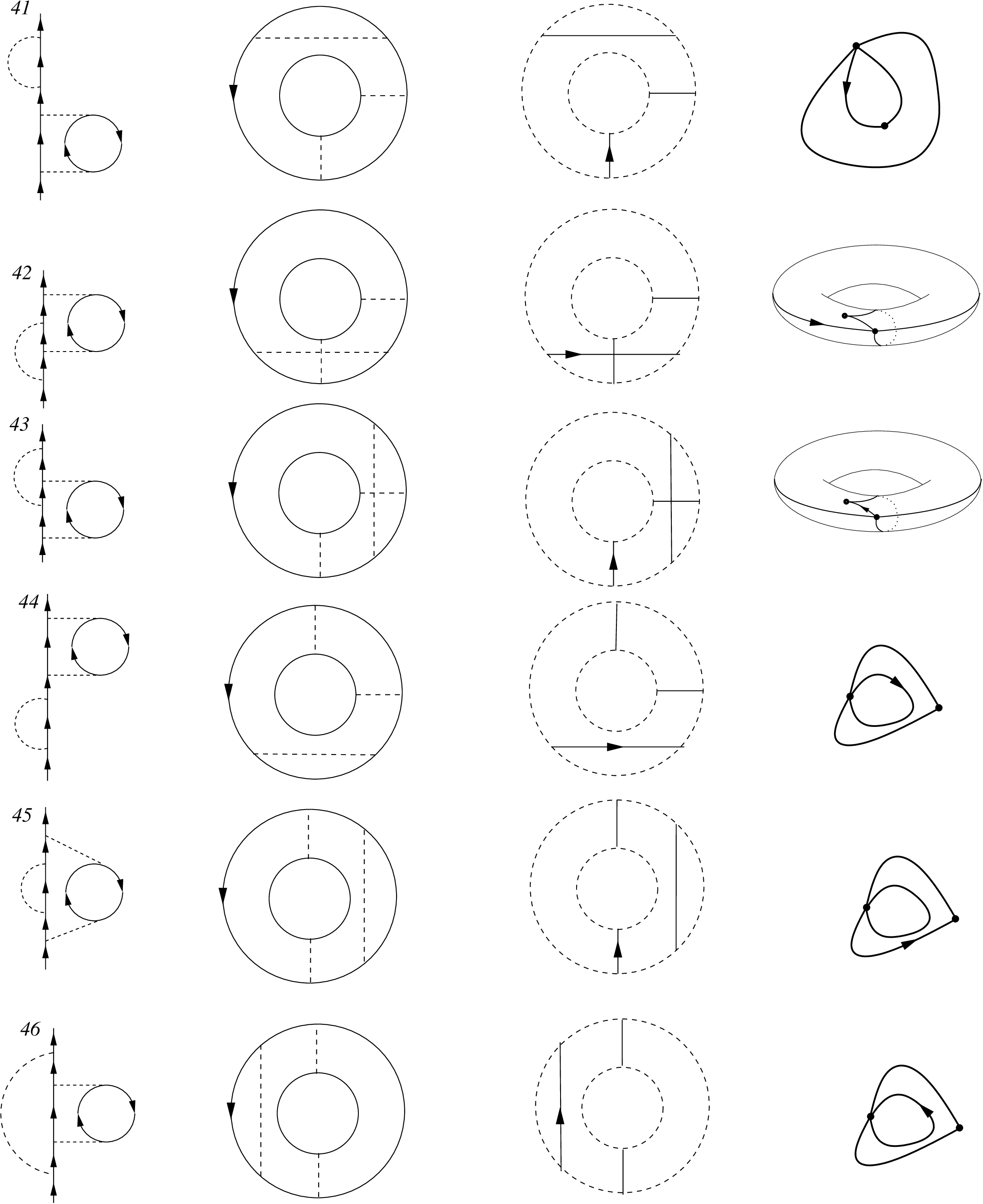}
\caption{Closed and quotient graphs of other third order diagrams and their 
related rooted maps.}\label{quot6}
\end{minipage}
\hspace{0.5cm}
\begin{minipage}[b]{0.47\textwidth}
\centering
\includegraphics[width=0.7\textwidth]{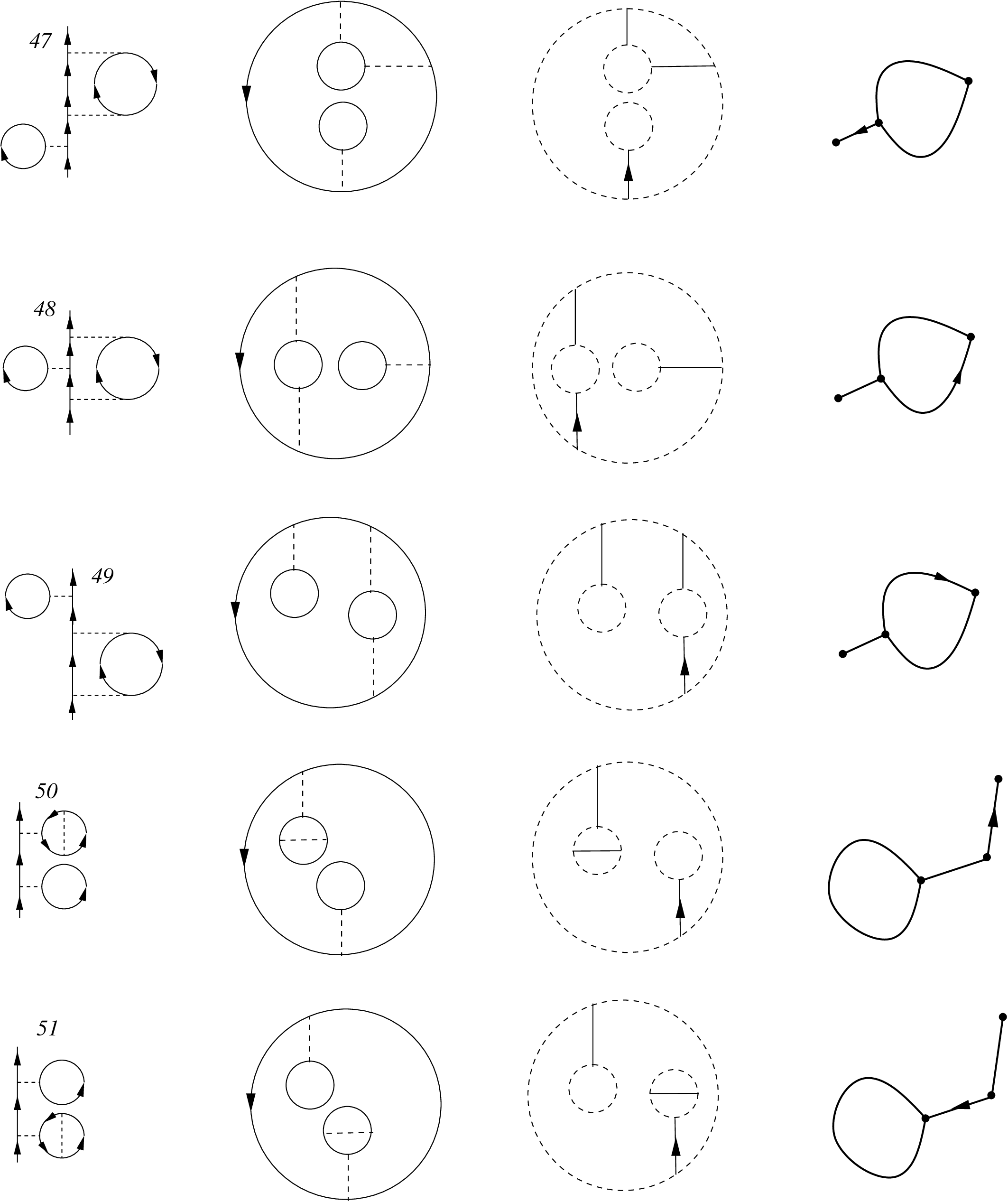}
\caption{Closed and quotient graphs of other third order diagrams and their 
related rooted maps.}\label{quot7}
\end{minipage}
\end{figure}
\begin{figure}
\begin{minipage}[b]{0.47\textwidth}
\centering
\includegraphics[width=0.7\textwidth]{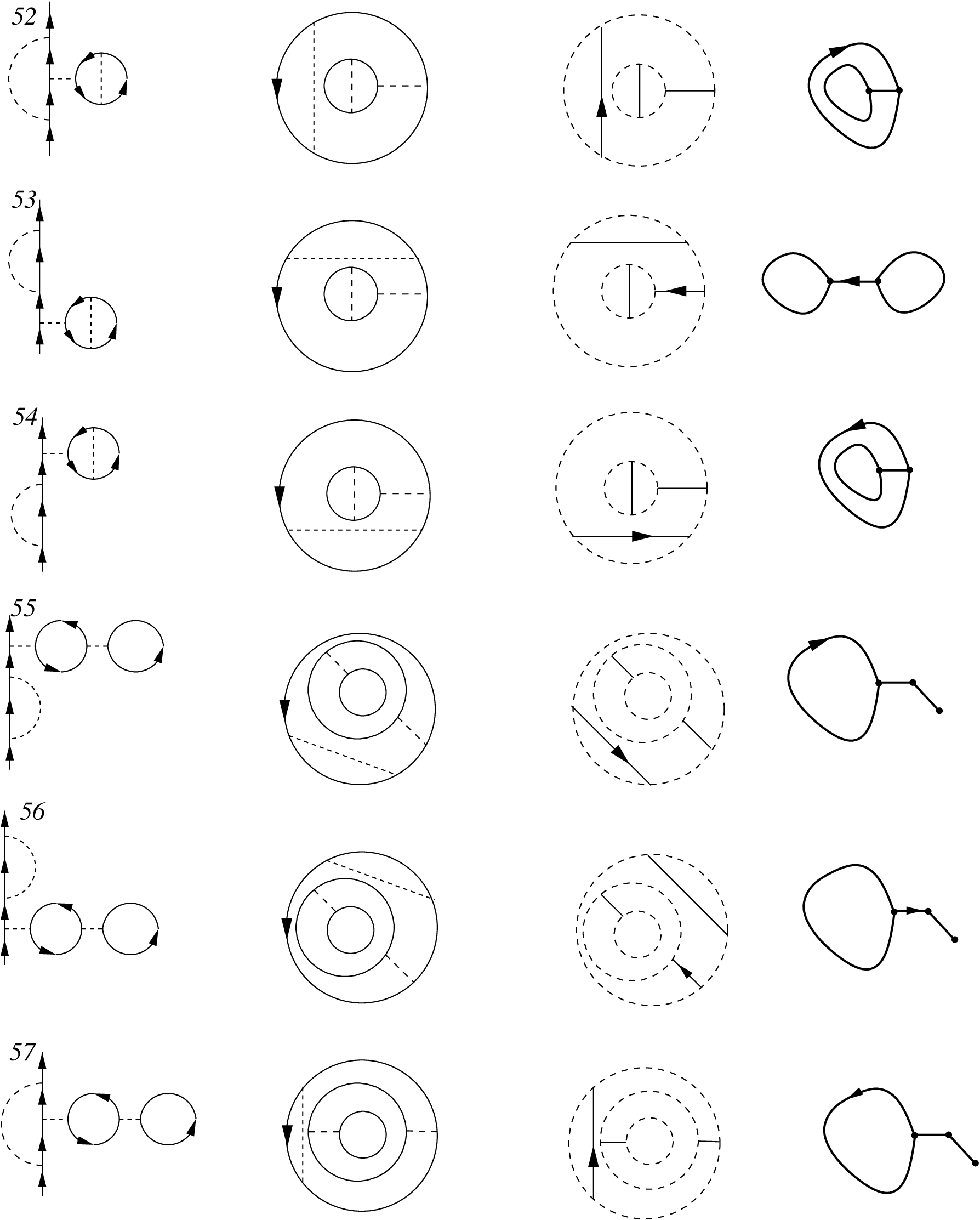}
\caption{Closed and quotient graphs of other third order diagrams and their 
related rooted maps.}\label{quot8}
\end{minipage}
\hspace{0.5cm}
\begin{minipage}[b]{0.47\textwidth}
\centering
\includegraphics[width=0.8\textwidth]{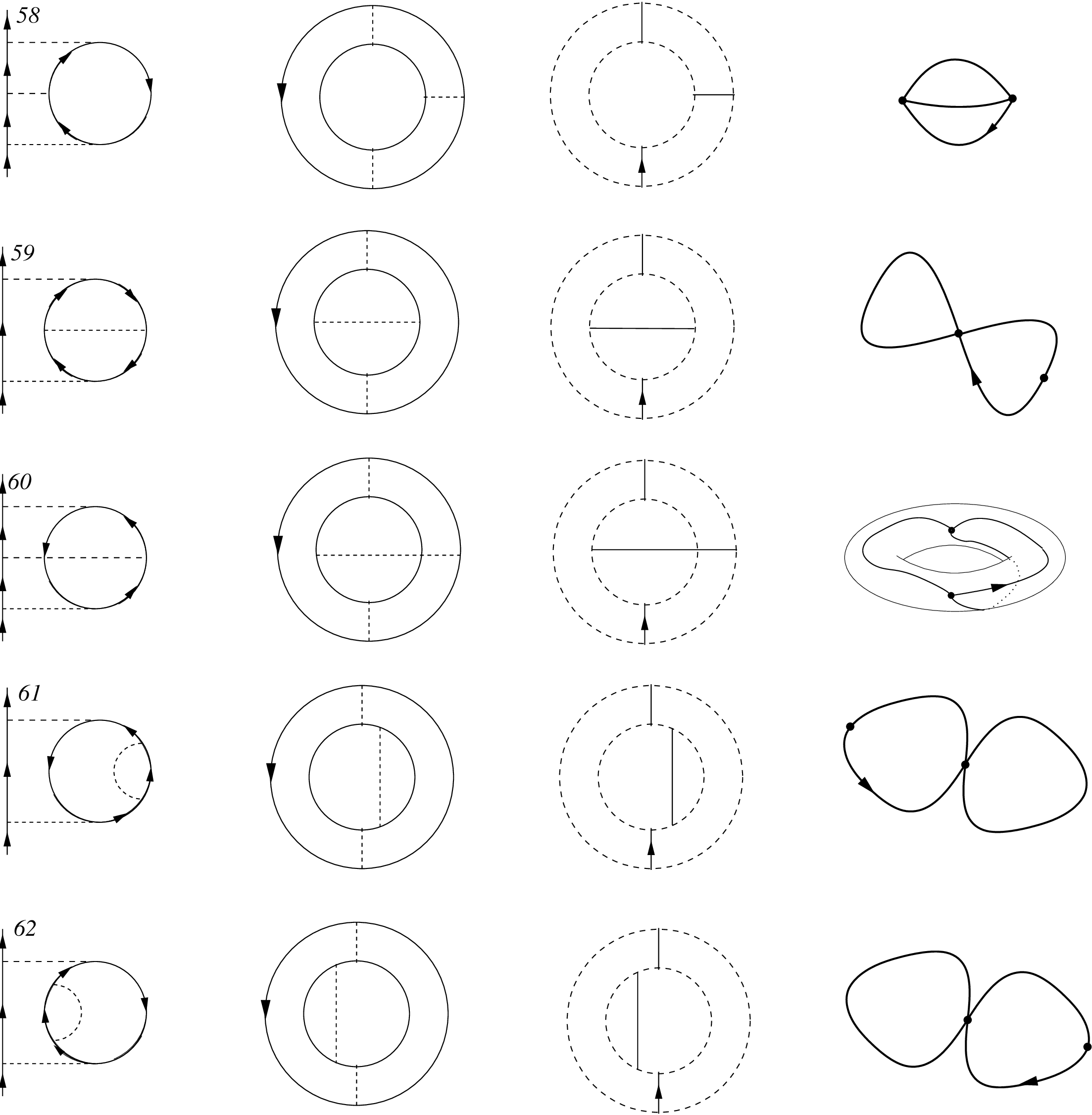}
\caption{Closed and quotient graphs of other third order diagrams and their 
related rooted maps.}\label{quot9}
\end{minipage}
\end{figure}
\begin{figure}
\begin{minipage}[b]{0.47\textwidth}
\centering
\includegraphics[width=0.8\textwidth]{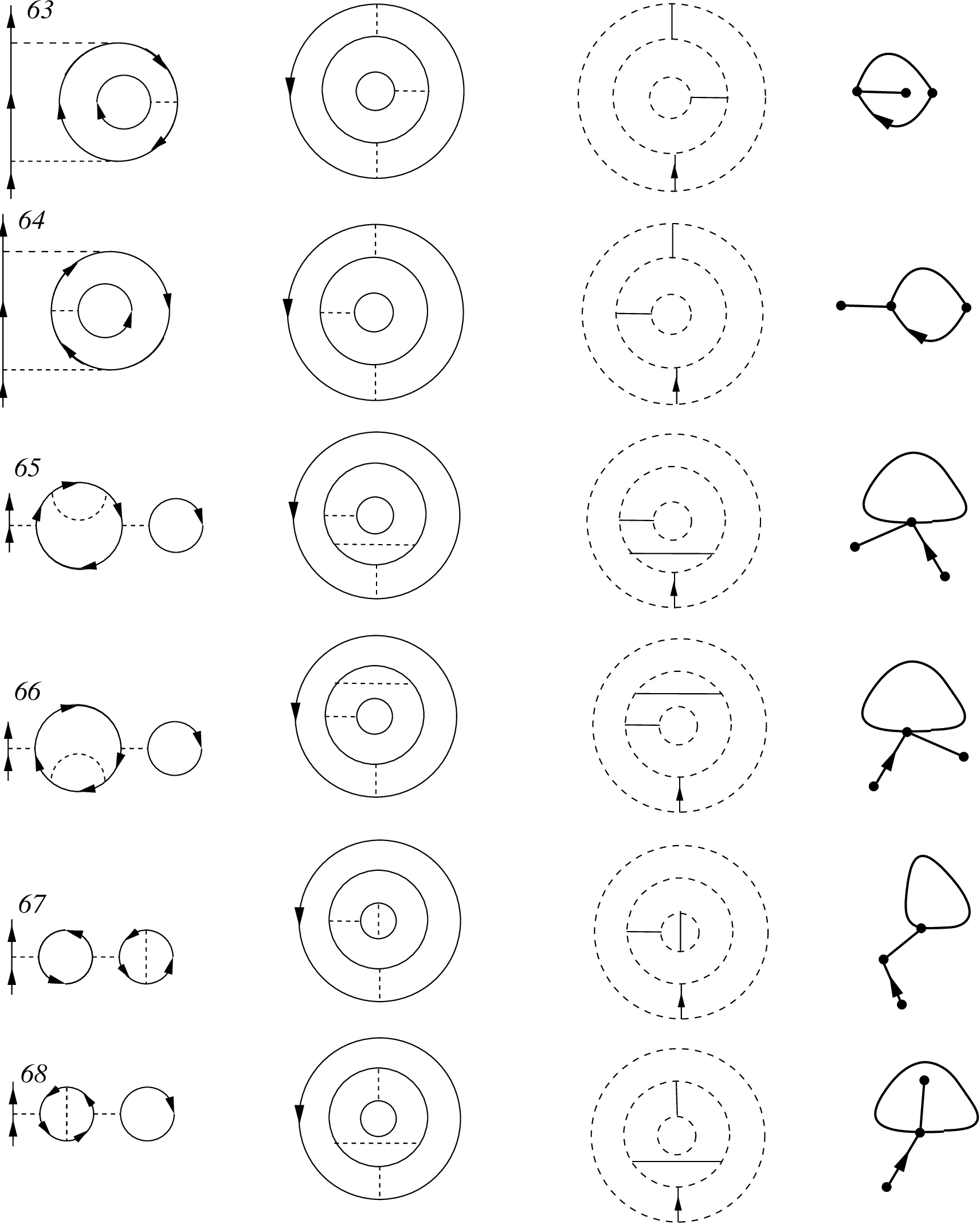}
\caption{Closed and quotient graphs of other third order diagrams and their 
related rooted maps.}\label{quot10}
\end{minipage}
\hspace{0.5cm}
\begin{minipage}[b]{0.47\textwidth}
\centering
\includegraphics[width=0.8\textwidth]{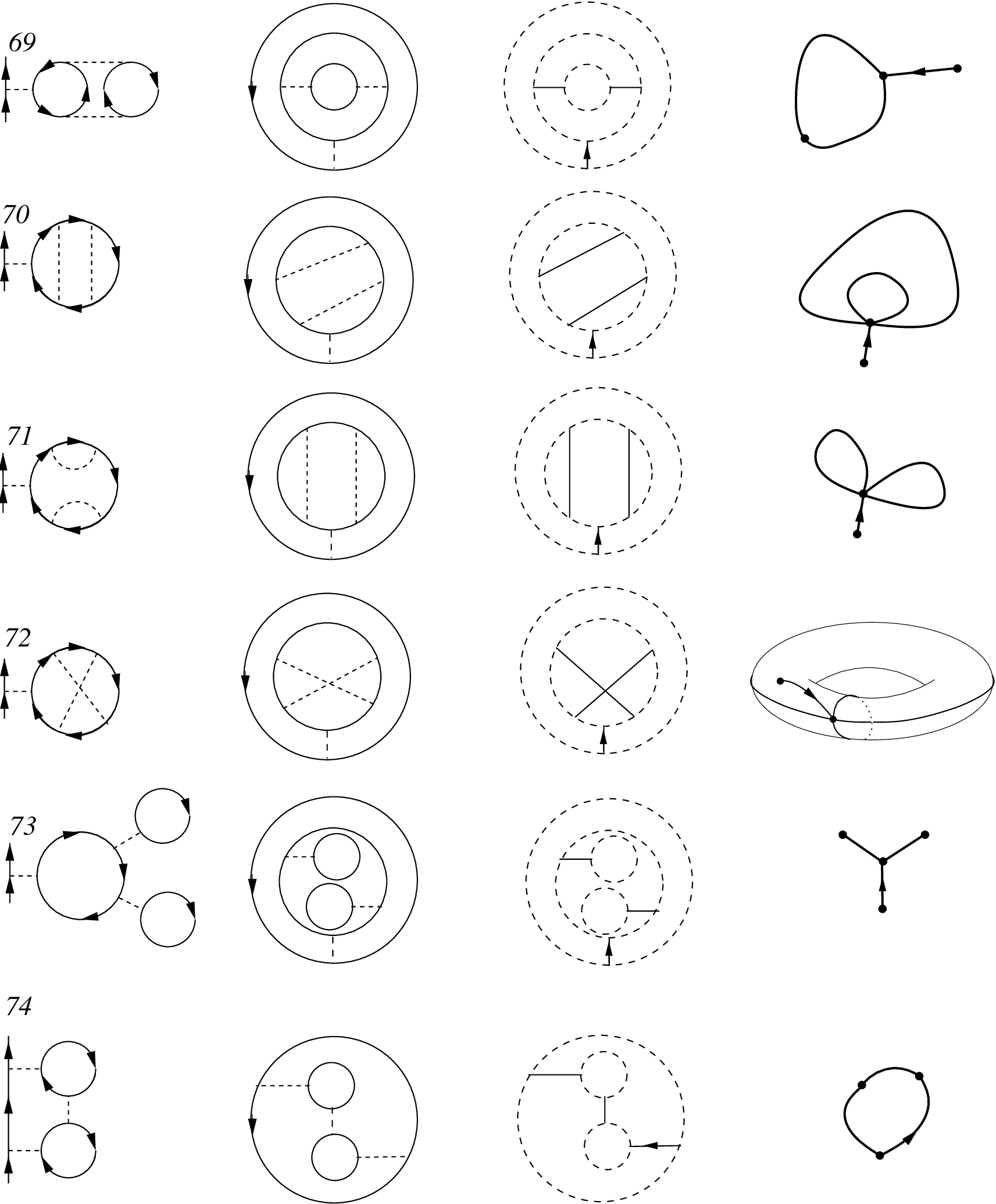}
\caption{Closed and quotient graphs of other third order diagrams and their 
related rooted maps.}\label{quot11}
\end{minipage}
\end{figure}
Let us analyze the number of vertices of this maps: we find that all of them 
have {\em one} vertex. But if we check our Table \ref{numeridiwalsh1}, we 
discover that on the sphere {\em there are just five  rooted maps with 3 edges 
and only one vertex}. This also ensures the validity of the the quotient 
procedure and hence the validity of the one-to-one correspondence with rooted maps on the sphere 
at this stage.\footnote{Notice that -- if the quotient procedure is right -- 
{\em there could not be} other maps with 3 edges, one vertex and genus 1, 
related to the five Feynman diagrams. In fact, if we check the Walsh work 
\cite{WI} on page 215, we do not find any other map with these features.}

We show all the quotient graphs of the remaining third order Feynman diagrams in 
Figs~\ref{quot1}, \ref{quot2}, \ref{quot3}, \ref{quot4}, \ref{quot5}, 
\ref{quot6}, \ref{quot7}, \ref{quot8}, \ref{quot9}, \ref{quot10}, \ref{quot11}.

\subsubsection{Remarks} Comparing the Tutte's rooted maps and the list of first, 
second and third order Feynman diagrams we can easily conclude that every 
Feynman diagram of the $n$-th order with $l$ loops -- and without crossing 
either between any interaction lines or between interaction and propagation 
lines -- is related to one (and only one) rooted map embedded in a sphere-like 
surface with $n$ edges and $l+1$ vertices. Through the quotient procedure, in 
fact, an additional loop is created (the one which contains the root edge) but 
all the propagation lines, after the shrinking procedure, become a point. For 
this reason the $l$ loops become $l+1$ {\em vertices} (which are actually 
points). Moreover, through the same procedure, the interaction lines of the 
original Feynman diagram are not modified so that they directly become the {\em 
edges of the map}. We can check the validity of this procedure by verifying that 
we have just obtained:
\begin{itemize}
\item 2 rooted maps on the sphere starting from the 2 Feynman diagrams at first 
order; among these 1-edge maps, we
have obtained: 1 rooted map with 1 vertex and one map with 2 vertices.
\item 9 rooted maps on the sphere starting from the 10 Feynman diagrams at 
second order; among these 2-edges maps
there are 2 maps with one vertex, 5 maps with 2 vertices and 2 maps with 3 
vertices.
\item 54 rooted maps with 3 edges on the sphere starting from the 74 Feynman 
diagrams at third order. 5 of them have
just 5 vertices, 22 of them have 2 vertices and other 22 have 3 vertices; 
finally we have obtained, among this class
of maps with 3 edges, 5 maps with 4 vertices.
\end{itemize}
This is a remarkable result because it is in perfect agreement with the works by 
Bender and Canfield  and with the ones by Walsh and Lehman (see Table \ref{numeridiwalsh1}).
\subsection{The genus of a Feynman diagram}
Finding the embedding of a graph was shown to be a NP-complete problem (see for 
instance Carsten \cite{NPC}). However, the existence of a precise association between 
Feynman diagrams and maps (based on the quotient procedure) strongly suggests that Feynman 
diagrams and rooted maps {\bf are isomorphic mathematical object}. Thus, it 
becomes natural to define {\em the genus of a Feynman diagram} as the genus of 
the related rooted map, i.e. the genus of the orientable surface in which the 
related rooted map is embedded. However, this definition is quite different from 
the one given in the work by J. S. Kang~\cite{Kang}, 
where the genus ($G$) of the Feynman diagram is defined as
\beq V-P+I=2-2G. \label{euler1} \eeq
\begin{table}[ht]
\caption{ Comparison between the Kang's definition of the genus $G$ 
of a Feynman diagram Eq. (\ref{euler1}) and the genus $g$ calculated by 
our definition Eq. (\ref{euler3}), with  regard to the Feynman diagrams 
(n.1, n.2, \ldots) represented in Fig. \ref{IIorder1} and Fig. \ref{IIorder2}.} 
\begin{center}
\begin{tabular}{ccc} 
{diagrams} & {vertices-propagators+loops$=2-2G$ } & 
{(loops+1)-edges+faces$=2-2g$} \\
\colrule \\
n.1 & $4-5+1=2-2G \rightarrow G=1$ & $2-2+2=2-2g \rightarrow g=0$ \\
n.2 & $4-5+1=2-2G \rightarrow G=1$ & $2-2+2=2-2g \rightarrow g=0$ \\
n.3 & $4-5+1=2-2G \rightarrow G=1$ & $2-2+2=2-2g \rightarrow g=0$ \\
n.4 & $4-5+1=2-2G \rightarrow G=1$ & $2-2+2=2-2g \rightarrow g=0$ \\
n.5 & $4-5+0=2-2G \rightarrow G=\frac{3}{2}$ & $1-2+3=2-2g \rightarrow g=0$ \\
n.6 & $4-5+2=2-2G \rightarrow G=\frac{1}{2}$ & $3-2+1=2-2g \rightarrow g=0$ \\
n.10 & $4-5+0=2-2G \rightarrow G=\frac{3}{2}$ & $1-2+1=2-2g \rightarrow g=1$ \\
\end{tabular}
\label{comparison}
\end{center}
\end{table}
Here $V$ is the number of vertices, $P$ the number of propagators and $I$ the 
number of the closed loops. 
This definition is followed by many 
authors like Gross, Mikhailov and Roiban \cite{Gross}, Nayak \cite{ Nayak} or 
Schwarz \cite{Sch}. We propose, instead, a definition which start from the well known Euler 
characteristic of a rooted map:
\beq v-e+f=2-2g, \label{euler2} \eeq
where $v$ is the number of the vertices of the map, $e$ the number of its edges 
and $f$ the number of its faces. Then
we substitute the number of vertices with $v=l+1$:
\beq (l+1)-e+f=2-2g, \label{euler3} \eeq
where $l$ is the number of loops in the associated Feynman diagram, $f$ and $g$ 
are respectively the number of faces and the number of holes  of the related 
rooted map (namely the genus of the orientable surface in proper sense). Thus 
the genus of Feynman diagrams provided by Eq.(\ref{euler3}) is similar to the one 
of formula (\ref{euler1}) {\em only} in the sense that both equations derive 
from the Euler-Poincar\'e formula. But the numbers $G$ and $g$ associated to the 
same diagram are usually different and only accidentally coincide, as reported 
in Table \ref{comparison}.
\section{Summary and Conclusions} \label{sec4}
In this work we have shown, up to third order in perturbation theory, the 
perfect agreement between the number of Feynman diagrams as a function 
of the perturbative order and the number of rooted maps on orientable surfaces 
as a function of the number of edges (and regardless to genus and to the number 
of vertices on the map).

This result has been obtained by establishing a graphical correspondence between Feynman
diagrams and rooted maps on the sphere and on oriented surfaces of higher genus. The quotient 
procedure presented here contains definite and unambiguous rules which allow to obtain 
the rooted map corresponding to a generic Feynman diagram.
In this connection it is worth pointing out that in the Appendix we define a simple but 
effective procedure for building Feynman diagrams at any order in perturbation theory 
from a purely graphical point of view.

A new definition of the {\em genus of a Feynman diagram} has also been given: 
the genus of a Feynman diagram is the number of holes
of the surface in which the corresponding rooted map is embedded. Hence
one might conjecture that Feynman diagrams and rooted maps are the same topological
object. The information about the physics is totally embodied in the number 
of vertices, edges, faces and holes of the embedding surface and their mutual relations.

It should be stressed that in this work we have considered Feynman diagrams entering into the 
perturbative expansion of the single-particle propagator within a many-body context. Other 
classes of diagrams, for example in the two-body propagator or the polarization propagator, 
as well as in the diagrammatic representation of the perturbative contributions to the ground/excited 
state energy of the system (Goldstone diagrams) are crucial for the correct determination of the
observables. Starting from the results of the present work it should be possible to calculate 
some specific class of diagrams such as ``ladder'', ``RPA'', and ``irreducible'' diagrams by 
means of rooted maps theory.  In particular it could be interesting a comparative approach to
 ``non-separable maps'', but this goes well beyond the scope of this paper.
\appendix
\section*{Appendix}

\subsubsection{Building Feynman diagrams} Let us build the Feynman diagrams for the 
single-particle Green's function at the $m$-th order in perturbation theory. For 
any given diagram there is an identical contribution from all similar diagrams 
that merely differ in the permutation of the space-time labels in the 
interaction Hamiltonian. 

In $m$-th order there are $m!$ ways of choosing the sequence of the interaction Hamiltonians by 
applying Wick's theorem \cite{Wick}. All of these terms give the same 
contribution to the Green's function, so that we can count each diagram just 
once and cancel the factor $\frac{1}{m!}$ in formula (\ref{newGreen}). Note that this 
result is true only for the connected diagrams, where the external points $x$ 
and $y$ are fixed. 

Let us now revisit the rules for building all the 
Feynman diagrams contributing to single-particle Green's function.

\begin{enumerate}
\item Draw all {\em topologically distinct} connected diagrams with $m$ 
interaction lines and  $2m+1$ oriented lines representing Green's functions. 
This procedure can be topologically simplified by observing that a Fermion line 
either closes on itself or runs continuously from $y$ to $x$. Each diagram 
represents all the $m!$ different
possibilities of ordering the set of space variables. If there is a problem 
concerning the precise meaning of topologically distinct diagrams, Wick's 
theorem can always be used to verify the enumeration.
\item Label each vertex with a four-dimensional space-time point.
\item Each solid line between the two points represents a free single-particle 
Green's function.
\item Each dashed line represents an interaction (more precisely its matrix 
element in spin-space).
\item Integrate over all internal space an time variables.
\item There is a spin matrix product along each continuous fermion line, 
including the potential at each vertex.
\item Affix a sign factor $(-1)^{l}$, where $l$ is the number of closed fermion 
loops in the diagram.\footnote{The overall sign of the various contributions 
appearing in 
the diagrams is determined as follows: every time a fermion line closes on 
itself, the term acquires an extra minus sign.}
\item To compute $G(x,y)$ assign a factor $\lp \frac{i}{\hbar}\rp^{m} =(-i)\lp- 
\frac{i}{\hbar}\rp^{m} i^{2m+1}$ to each $m$-th order term.
\end{enumerate}
\begin{figure}
\begin{minipage}[b]{0.47\textwidth}
\centering
\includegraphics[width=0.4\textwidth]{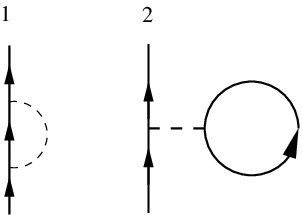}
\caption{First order diagrams.}
\label{1}
\vspace{0.1cm}
\end{minipage}
\hspace{0.5cm}
\begin{minipage}[b]{0.47\textwidth}
\centering
\includegraphics[width=0.95\textwidth]{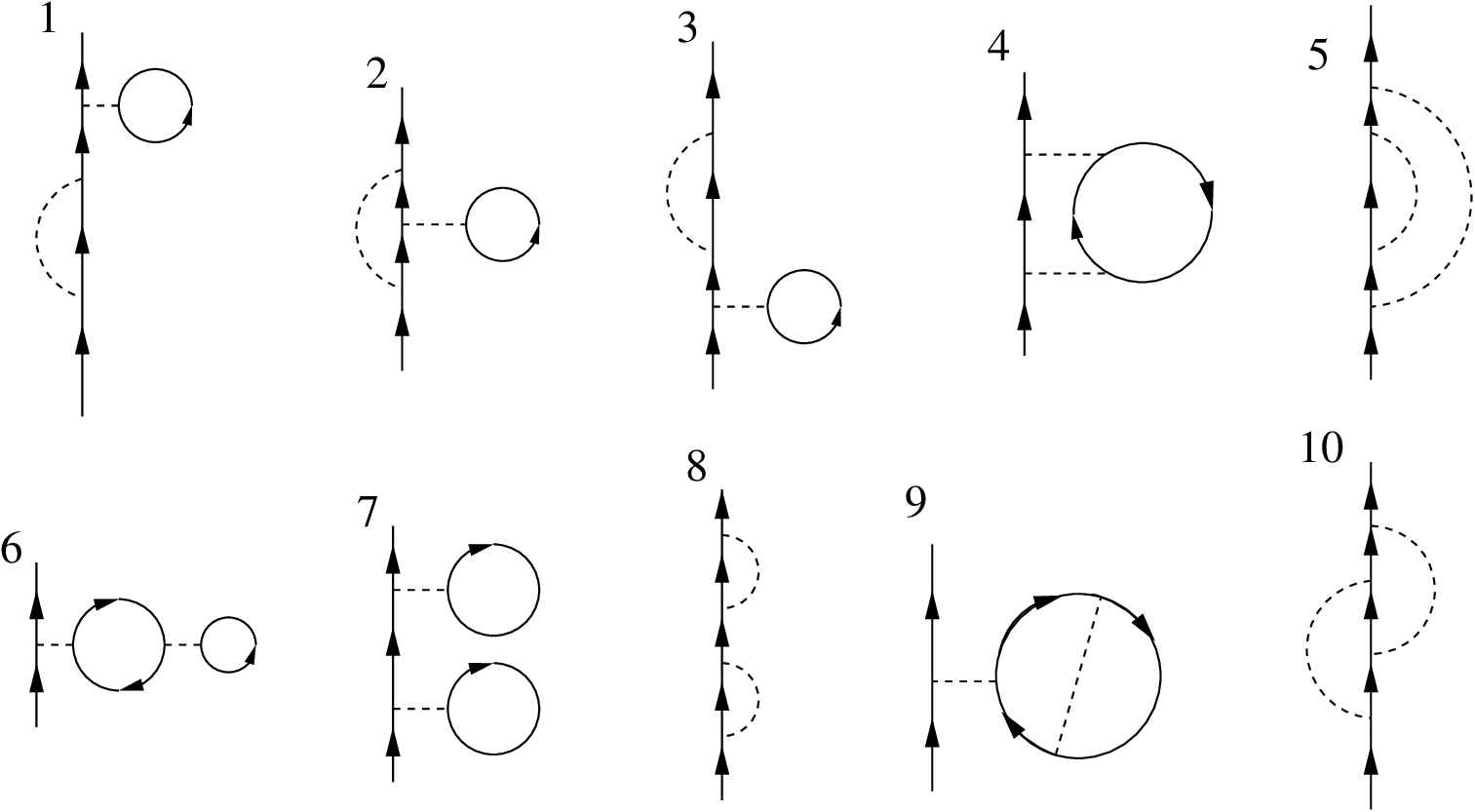}
\caption{Second order Feynman diagrams  
\cite{Wale}.} \label{2}
\end{minipage}
\end{figure}

The foregoing statements provide a unique prescription for drawing and evaluating all Feynman 
diagrams that contribute to the Green's function in coordinate space. Each 
diagram corresponds to an analytic expression that can be now written down 
explicitly with the Feynman rules. The calculation of Green's function becomes 
then a relatively automatic although non-trivial process.
\subsubsection{First and second order diagrams}
As an example of the Feynman rules, we show  the complete first order 
contribution to the Green's function in Fig.\ref{1}.

The corresponding second order contribution requires more work and, according to 
 Fetter and Walecka~\cite{Wale}, we can assert that there are 10 second order 
Feynman diagrams (see Fig.\ref{2}).

Historically speaking, Feynman diagrams are not the first graphical approach to 
a complex physics problem. First efforts were made in the integration of 
dynamical systems in the realm of thermodynamics and statistical mechanics. An 
interesting example of this approach is the one developed by Ursell and Mayer 
\cite{Ur,May} in order to determine the partition function of an interacting 
gas, known as the {\em cluster expansion}.
\subsubsection{Third order diagrams} Even though the number of Feynman diagrams for 
the electron propagator has been found out many years ago, it seems 
that a graphical enumeration of third order diagrams (or beyond) has never been 
done. It may seem a rather boring and useless work drawing all these diagrams. 
And surely it is, if we merely consider the physics point of view. Nevertheless, in 
order to make a comparison with rooted maps, which are essential in 
characterizing the topological properties of orientable surfaces, the graphical identification and 
enumeration of the diagrams may be an useful tool. 
Thus we looked both to the {\em number} of Feynman diagrams at 
third order and to the {em structure} of each diagram. For this 
purpose, we have devised an easy procedure for drawing all topologically distinct 
Feynman diagrams at {\em any} order. First of all we built the 10 second order 
diagrams starting from the ``shell'' and the ``tadpole'' diagrams of first 
order. This way we can enumerate the first three numbers of the puzzling 
``quantum many-body theory integer sequence'' (perturbation order N versus 
number of quantum many-body theory diagrams).

\begin{figure}
\begin{center}
\includegraphics[width=.55\textwidth]{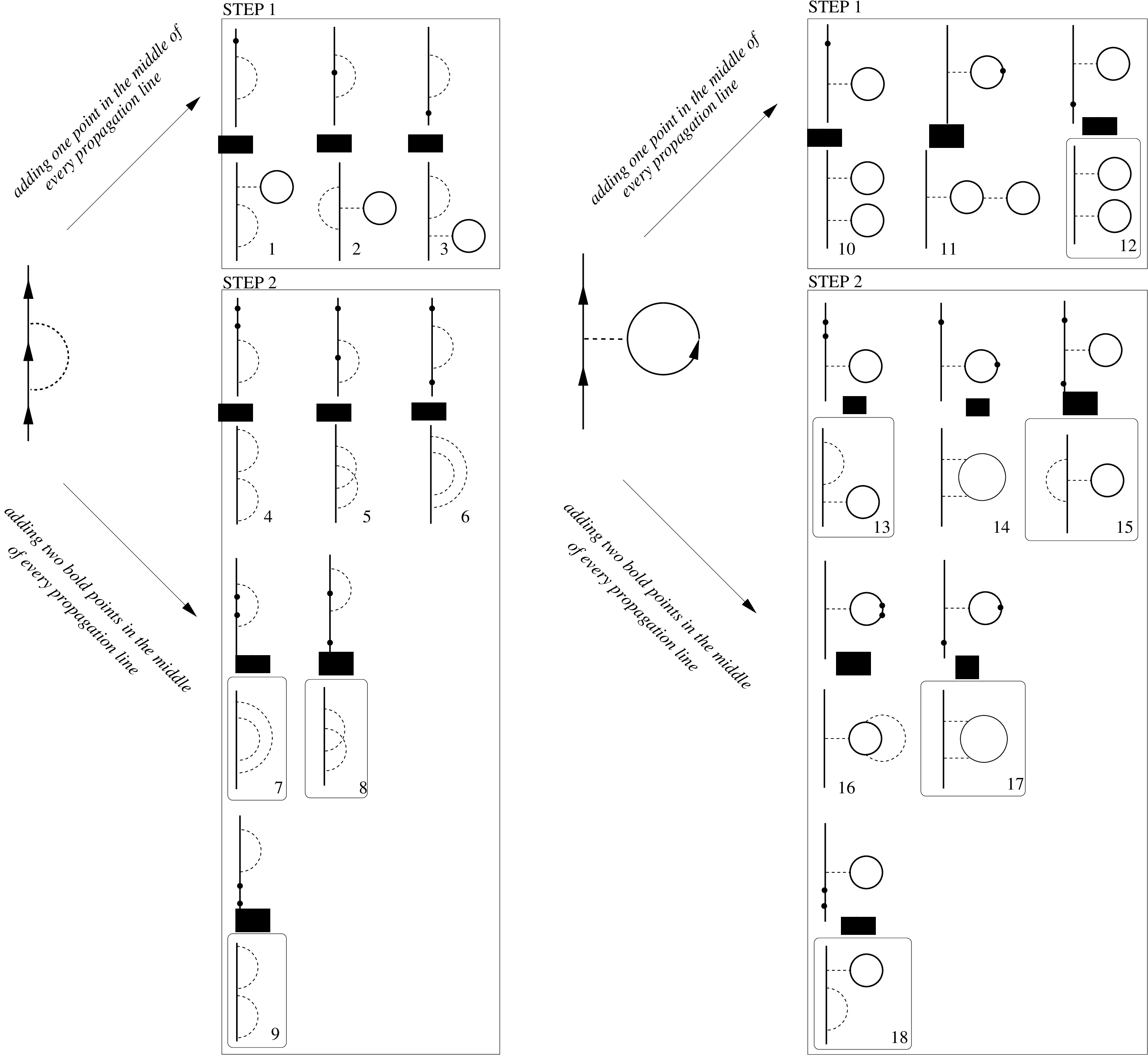}
\caption{ Examples of the graphical construction for the second order Green's function diagrams, 
starting from the two diagrams at first order: ``Shell'' and 
``Tadpole''. In step 1 we put {\em one bold} point in the middle of a 
propagation line; then we add a {\em tadpole} where we have just put that 
point. In step 2 we add, analogously, {\em two} bold points and connect them 
with a dashed line representing a new {\em shell} interaction. 
Obviously these points must be put in every 
available position. We thus obtain 18 diagrams but it is easy to observe that 
8 of them (the ones shown inside rectangular boxes) appear twice and must be 
discarded as topologically equivalent to other diagrams (in agreement with Feynman 
rule number 1).} \label{building}
\end{center}
\end{figure}

We show the building procedure (see Fig.\ref{building}) applied to first order 
diagrams in such a way as to obtain the (already known) second order ones and 
in general to show how to apply it for higher orders:
\begin{enumerate}
\item Consider the ``shell'' diagram.
\item We draw {\em one} bold point in the middle of a propagation line: it will 
become the tadpole tail of a new second order diagram; of course we can draw a 
bold point in three different positions\footnote{Often we refer to the set of 
segments representing Green functions and connecting the initial and final 
vertices straightforwardly as the
{\em root line}}, so that we should built three new different second order 
quantum many-body theory diagrams, starting from the first-order shell one.
\item Then we have to put {\em two} bold points on any propagation line, and 
they will become the shell extremes of a new second order diagram.
\item Repeat steps 1,2 and 3 starting with the ``tadpole'' diagram.
\end{enumerate}
The procedure is the same for higher orders: to obtain the $m$th-order diagrams 
 one has to apply it to every diagram of the $(m-1)$th-order. The result of our
building procedure applied to the first order diagram of Fig.\ref{1} are the 10 
 second order diagrams of Fig.\ref{2}. For the third order, we have to start from 
the 10 distinct diagrams of the second order, and so on...

This way, we can build all Feynman diagrams: we notice that {\em the extreme of an 
interaction line can only be attached to a propagation line, hence to the ``points'' 
systematically added in the lower order diagram}. 
Obviously similar diagrams may arise from this procedure: 
therefore at the end of the process, we have to discard 
{\em topologically equivalent} diagrams, which are such if there exist a continuous transformation 
in the plane between each part of the two diagrams (see Fig.\ref{building}).
The physical meaning of this rules directly stems from the Wick's Theorem 
\cite{Wick} applied to the propagator. 

\begin{figure}
\begin{minipage}[b]{0.47\textwidth}
\centering
\includegraphics[width=0.95\textwidth]{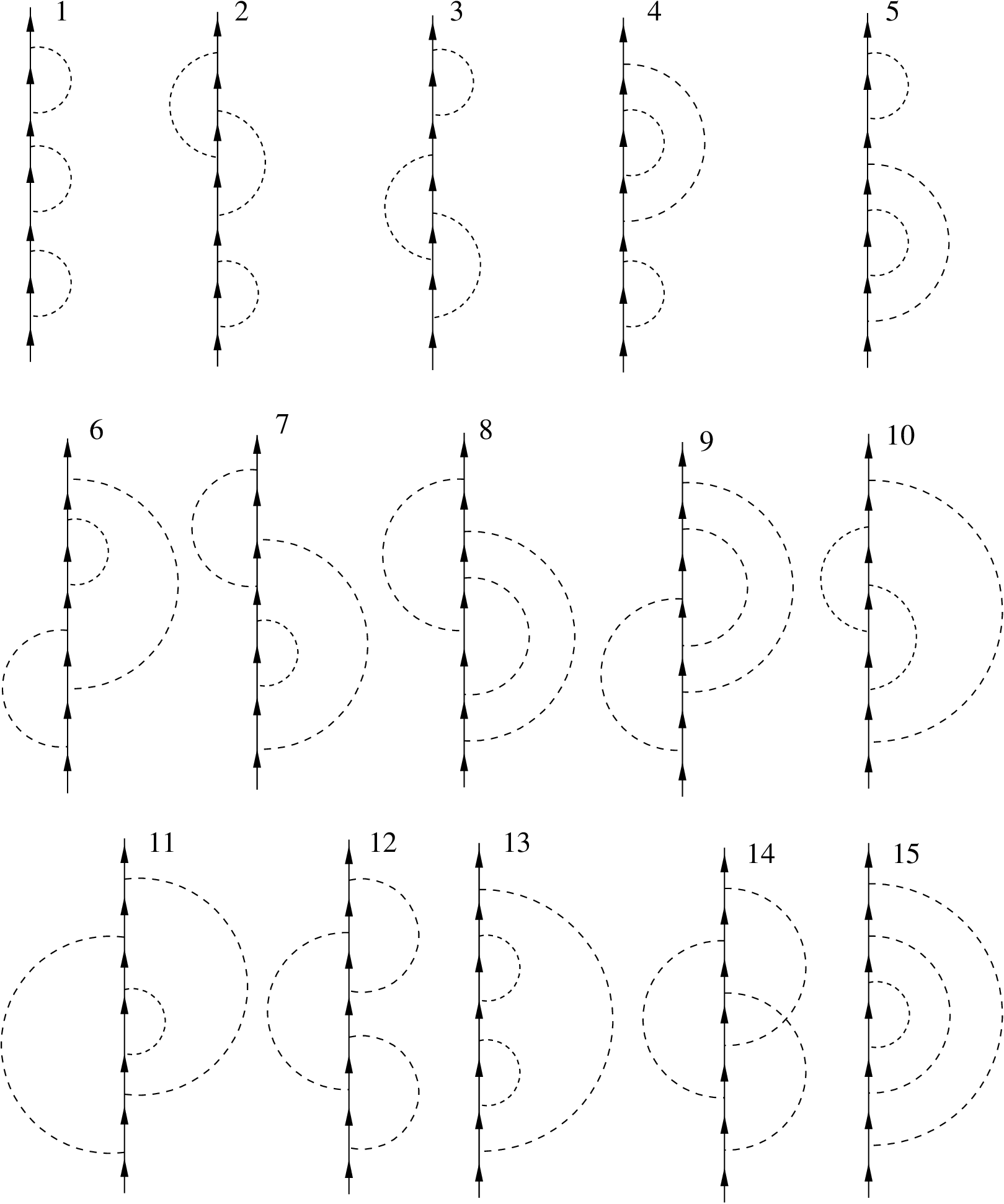}
\caption{ ``Shells'' Feynman diagrams of third order. Interaction vertices are 
added only on the root of
shells diagrams of preceding order. These diagrams are simply computed by the 
Touchard formula: if $m$ is the
perturbative order (i.e. number of shells), the number of shells-diagrams is 
$(2m-1)!!$.} \label{shells}
\end{minipage}
\hspace{0.5cm}
\begin{minipage}[b]{0.47\textwidth}
\centering
\includegraphics[width=0.95\textwidth]{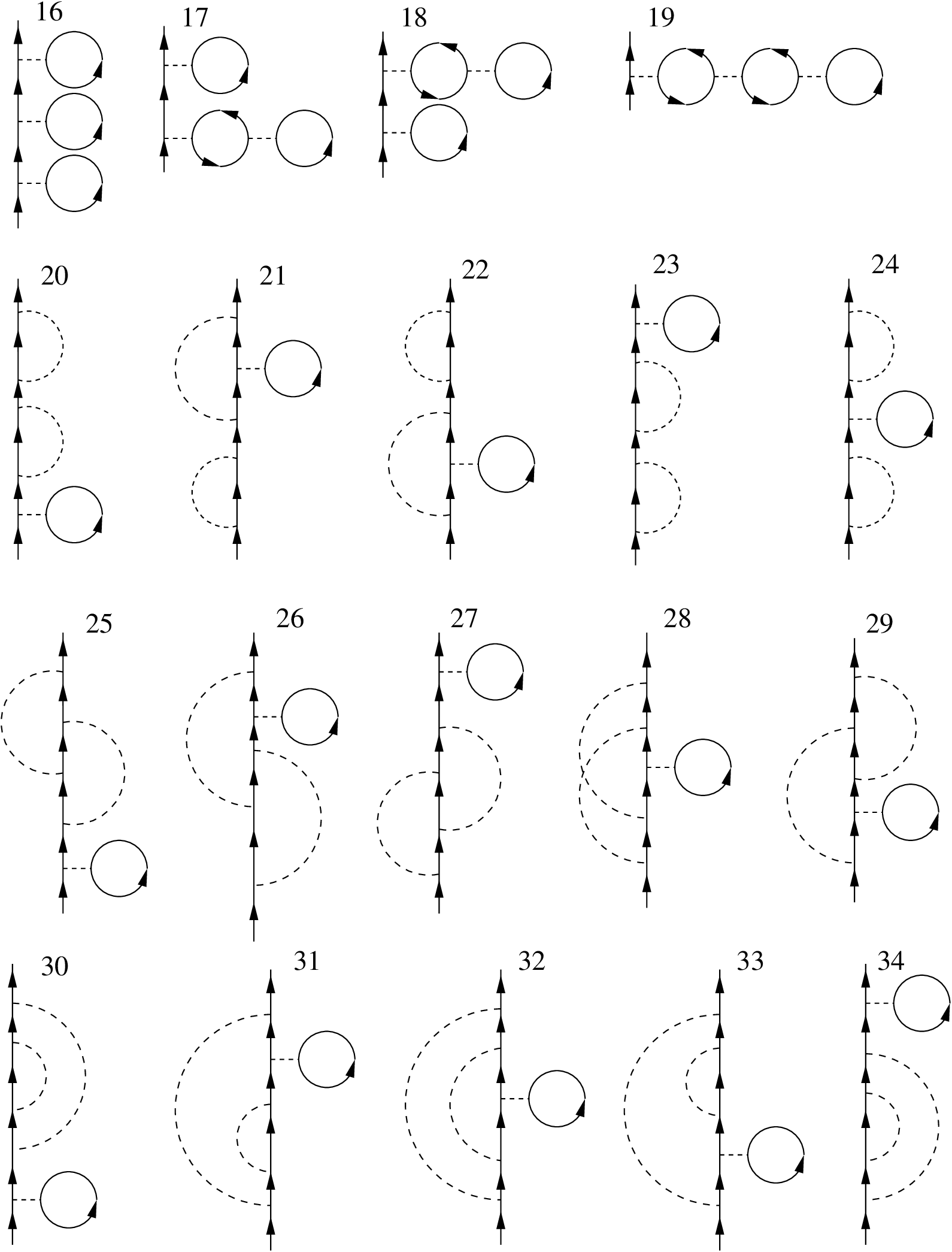}
\caption{ (a) Third order Feynman diagrams. ({\em continued})} \label{74a}
\vspace{1.2cm}
\end{minipage}
\end{figure}

We now show the results of the above established procedure for building all third order diagrams. 
As expected, we have obtained 74 different diagrams: 15 of them -- we 
loosely refer to them as to ``shells'' diagrams (shown in Fig.\ref{shells}), -- 
can be enumerated by a very easy formula since  
they contain only shells and no tadpoles\footnote{In this class of graphs, 
interaction vertices lie on the same propagation line directly connecting the 
initial and final vertices of the many-body Green's function diagram.}. At  
the $m$-th perturbative order their number is:

$$F_{shell}(m)=(2m-1)!!=1 \cdot 3 \cdot 5 \cdot 7 \textrm{\ldots}. $$

This formula was extracted by Touchard in an old strictly mathematical work 
\cite{Touch}, and recently rediscovered by several physicists among whom 
Kuchinskii \cite{Ku}.
\begin{figure}
\begin{minipage}[b]{0.47\textwidth}
\centering
\includegraphics[width=0.95\textwidth]{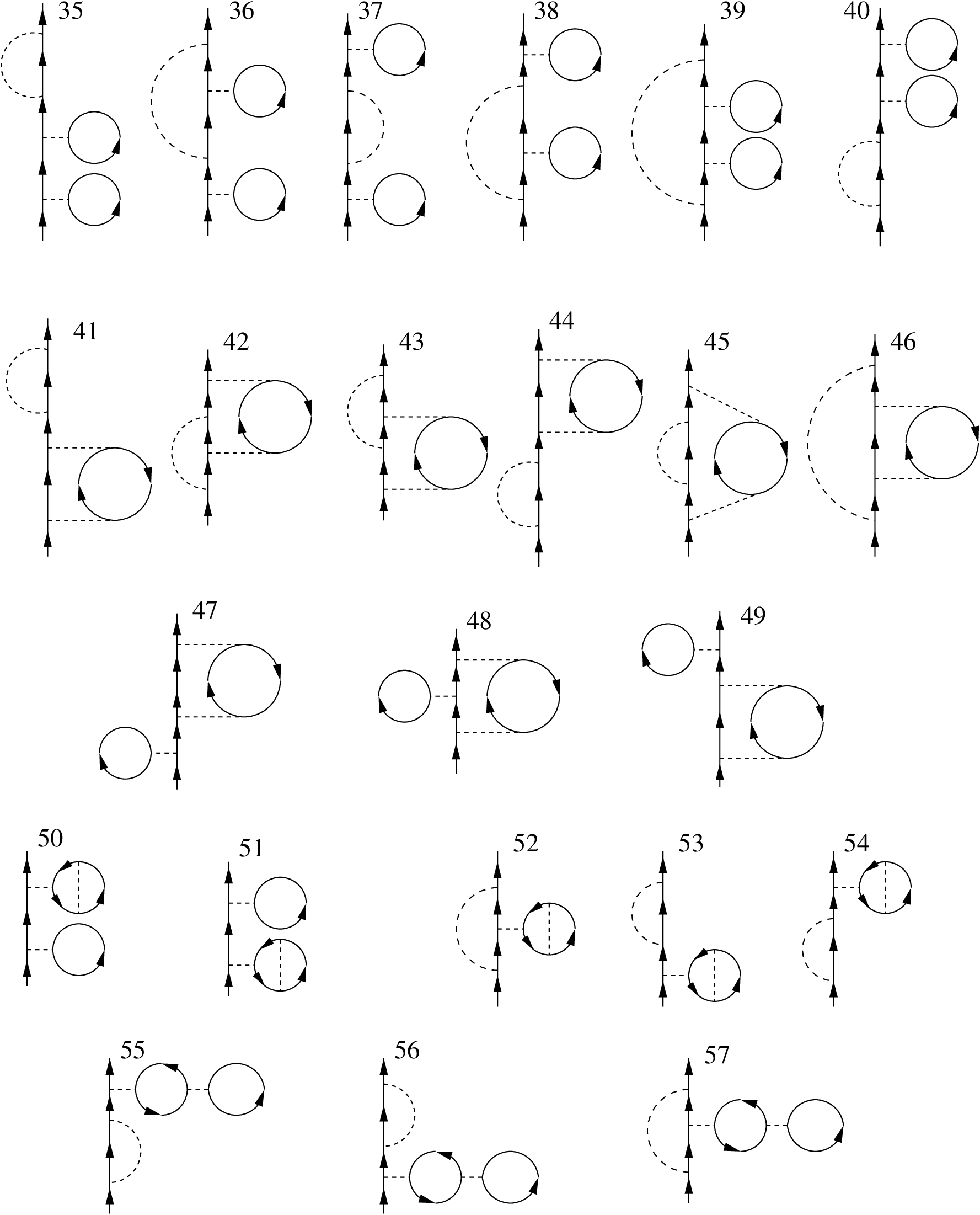}
\caption{ (b) Third order Feynman diagrams ({\em continued}).} \label{74b}
\end{minipage}
\hspace{0.5cm}
\begin{minipage}[b]{0.47\textwidth}
\centering
\includegraphics[width=0.95\textwidth]{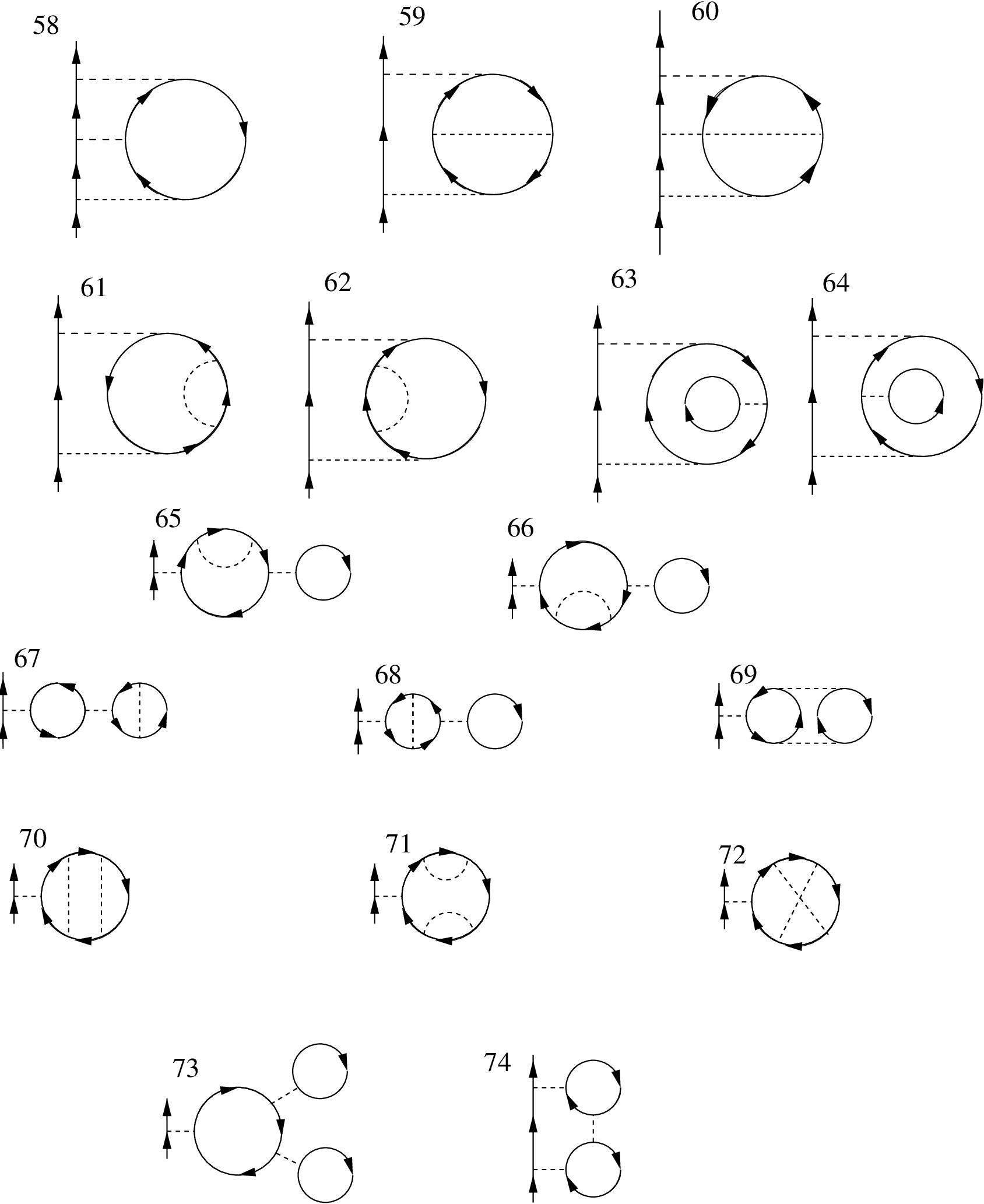}
\caption{ (c) Third order Feynman diagrams ({\em continued}).} \label{17}
\end{minipage}
\end{figure}

Checking all topologically equivalent diagrams, we found that there exist 42 
additional diagrams which are obtained by adding a shell or a tadpole in every 
available position right in the root line (see Figs~\ref{74a} and \ref{74b}). 
Moreover there are 18 diagrams derived from second order diagrams by adding a 
tadpole (or a shell) onto a tadpole -- or a shell between the root line and a 
tadpole (Fig.\ref{17}). 
The complete set of the single-particle Green's function Feynman diagrams at third order 
(in addition to the ones in Fig.\ref{shells}) is 
illustrated in Figs~\ref{74a}, \ref{74b} and \ref{17}.


\begin{thebibliography}{99}

\bibitem{floer1} A. Floer, Comm. Pure Appl. Math. 41  (1988) 775–813.

\bibitem{floer2} A. Floer, J. Diff. Geom. 30 (1989) 202–221.

\bibitem{floer3} A. Floer, Comm. Math. Phys. 118 (1988) 215–240.

\bibitem{fukaya1} K. Fukaya, Geometric Topology: 1993 Georgia International 
Topology Conference, August 2-13, 1993, University of Georgia, Athens, Georgia 
red. William Hilal Kazez.

\bibitem{fukaya2} K. Fukaya, Commun. Math. Phys. 181 (1996) 37-90.

\bibitem{fukaya3} K. Fukaya, Y.-G. Oh, H. Ohta, K. Ono,  arXiv:0912.2646v2.

\bibitem{fukaya4} K. Fukaya, Y.-G. Oh, H. Ohta, K. Ono, Duke. Math. J. 151 (2010) 23-174.

\bibitem{sch1} A. Goupi, G. Schaeffer,   Europ. J. Combinatorics 19(7)  (1998) 819-834.

\bibitem{sch2} G. Schaeffer and B. Jacquard,  J. Combin. Theory Ser. A83(1) (1998) 1-20.

\bibitem{sch3} M. Bousquet-Melou and G. Schaeffer, Adv. in Applied Math. 24 (2000) 337-368.

\bibitem{sch4} D. Poulalhon and G. Schaeffer,  Theoretical Computer Science 307(2) (2003) 385-401.

\bibitem{sch5} N. Bonichon, C. Gavoille, N. Hanusse, D. Poulalhon and G. 
Schaeffer, Graphs and Combinatorics 22(2) (2006) 185-202.

\bibitem{sch6} E. Fusy, D. Poulalhon and G. Schaeffer, European Journal of 
Combinatorics 30(7) (2009) 1646-1658.

\bibitem{ati} M. Atiyah,  Proceedings of Symposia in Pure Mathematics 48 (1988) 285–299.

\bibitem{witt1} E. Witten, J. Differential Geometry 17 (1982) 661–692.

\bibitem{witt2} F. Cachazo, P. Svrcekb, E. Witten, JHEP09(2004)006.

\bibitem{thooft} G. ’t Hooft, arXiv:gr-qc/9310026v2.

\bibitem{dewolf1} G. Ivanyos, H. Klauck, T. Lee, M. Santha, R. de Wolf, arXiv:1204.4596v1.

\bibitem{DiFr} P. Di Francesco, P. H. Ginsparg and J. Zinn-Justin, Phys. Rep. 254 (1995) 1.

\bibitem{R&K} P.J. Rossky, M. Karplus, J. Chem. Phys. 64 (1976) 1596.

\bibitem{Ku} E. Z. Kuchinskii, M. V. Sadovskii, JETP 86 (1998) 367.

\bibitem{Kl} H. Kleinert, A. Pelster, B. Kasteing, M. Bachmann,  Phys. Rev. E62 (2000) 1537-1559.

\bibitem{Ridd} R. J. Riddel Jr.,  Phys. Rev. 91 (1953) 1243.

\bibitem{BrI} C. Brouder, Euro. Phys. J. C4 (2002) 1-45.

\bibitem{BrII} C. Brouder,  Eur. Phy. J. C12 (2000) 535-549.

\bibitem{BrIII} C. Brouder, Eur. Phy. J. C19 (2001) 715-741.

\bibitem{C} P. Cvitanovi\'c, B. Lautrup, R. B. Pearson, Phys. Rev. D18 (1978) 1939-55.

\bibitem{Arques-rooted} D. Arqu\`es, J.-F. B\'eraud,  Disc. Math. 215 (2000) 1-12.

\bibitem{Arques-torus} D. Arqu\`es,  J. combin. Theory Ser. B43 (1987) 253-274.

\bibitem{furry} H. Furry, Phys. Rev. 51 (1939) 125.

\bibitem{Tutte1} W. T. Tutte, Canad. J. Math. 14 (1962) 708-722.

\bibitem{Tutte2} W. T. Tutte, Canad. J. Math.  15 (1963) 249-271.

\bibitem{Tutte3} W. T. Tutte, Bull. Amer. Math. Soc.  74 (1968) 64-74.

\bibitem{Brown1} W. G. Brown, {\em Enumerative problems of linear graph theory, 
Ph D. Thesis}, University of Toronto (1963).

\bibitem{Brown2} W. G. Brown, Mem. Amer. Math. Soc.  65 (1966) 1-42.

\bibitem{Brown3} W. G. Brown, W. T. Tutte,  Canad. J. Math. 16 (1964) 572-577.

\bibitem{WI} T. R. Walsh, A. B. Lehman, J. Comb. Theory Ser.B13 (1972) 192-218.

\bibitem{WII} T. R. Walsh, A. B. Lehman, J. Combin. Theory Ser. B13 (1972) 122-141.

\bibitem{WIII} T. R. Walsh, A. B. Lehman, J. Combin. Theory Ser. B18 (1975) 222-259.

\bibitem{Cai} S. S. Cairns, ``Introductory Topology'', Ronald Press, New York, 1961.

\bibitem{BI} E. A. Bender, E. R. Canfield, SIAM J. Disc. Math. 7  (1994) 9-15.

\bibitem{BII} E. A. Bender, E. R. Canfield, J. Comb. Th. A43 (1986) 244.

\bibitem{BIII} E. A. Bender, E. R. Canfield, R. W. Robinson, Canad. Math. Bull. 31 (1988) 257-271.

\bibitem{BIV} E. A. Bender, E. R. Canfield, J. Comb. Th. B53 (1991) 293-299.

\bibitem{Touch} J. Touchard,  Canad. J. Math.  4 (1952) 2-25.

\bibitem{Cou} B. Courcelle, V. Dussuax,  The electronic journal of combinatorics 9 (2002) n. R40.

\bibitem{KI} M. A. Krikun, V. A. Malyshev, Discrete Mathematics and Applications 11 (2001) 105-212.

\bibitem{KII} M. A. Krikun, V. A. Malyshev, ``Trends in Mathematics. Mathematics 
and Computer Science'', Ed. D. Gardy, A. Mokkadem. BirkHauser, 2002.

\bibitem{M} V. A. Malyshev, A. Vershik, {\em Combinatorics and Probability of 
Maps} in ``Asymptotic Combinatorics with Applications to Mathematical Physics'', 
NATO Science Series, 77, Kluwer, 2002, 71-95.

\bibitem{SI} G. Schaeffer,  The Electronic Journal of Combinatorics  4 (1997) n. R20.

\bibitem{SII} G. Schaeffer, D. Poulalhon, {\em A note on bipartite Eulerian planar maps}. To appear.

\bibitem{SIII} G. Schaeffer, {\em Enum\'eration de cartes planaires 
g\'en\'eralis\'ees}, ``S\'eminaire Algorithmique'' held in the framework of the 
``Laboratoire d'informatique de l'\'Ecole Polytechnique'', november 9th, 1995.

\bibitem{J} D. M. Jackson, T. I. Visentin, ``An atlas of the smaller maps in 
orientable and nonorientable surfaces'', Chapman \& Hall (2000).

\bibitem{Y} L. Yanpei, ``Enumerative Theory of Maps'', Kluwer Academic 
Publishers, Dordrecht. Co-publication with Science Press Beijing Hardbound, ISBN 
0-7923-5599-7 June 2000.

\bibitem{dodson} C.T. Dodson, P.E. Parker,``A user's guide to algebraic 
topology'', Mathematics and its Applications, 387. Kluwer Academic Publishers 
Group, Dordrecht, 1997.

\bibitem{Wale} Alexander L. Fetter, John D. Walecka, ``Quantum theory of 
many-particle systems'', McGraw-Hill Publishing Company (1971).

\bibitem{NPC} T. Carsten ``The graph genus problem is NP-complete'', Journal of 
Algorithms 10 (4): 568-576 (1989).

\bibitem{Kang} J. S. Kang,  Phys. Rev. D14 (1976) 1587-1601.

\bibitem{Gross} D. J. Gross, A. Mikhailov, R. Roiban, JHEP 0305 (2003) 025.

\bibitem{Nayak} C. Nayak, ``Many Body Physics'', University of California, Los Angeles, January 1999.

\bibitem{Sch} J. H. Schwarz, arXiv:astro-ph/0304507v1.

\bibitem{Wick} G. C. Wick, Phys. Rev. 80 (1950) 268.

\bibitem{Ur} H. D. Ursell, Proc. Cambridge Phil. Soc 23, 685 (1927).

\bibitem{May} J. E. Mayer and M. G. Mayer, ``Statistical Mechanics'', John Wiley 
and Sons, New York (1941).

\end{thebibliography}
\end{document}